\documentclass[12pt]{article}

\usepackage[margin=1in]{geometry}
\usepackage[T1]{fontenc}
\usepackage{setspace}
\usepackage{authblk} 
\usepackage{hyperref}
\usepackage{microtype} 
\usepackage{graphicx}
\graphicspath{{Fig/}}
\usepackage{subcaption}
\usepackage{booktabs,color} 
\usepackage{amsmath,amssymb,amsfonts,amsthm}
\usepackage{bm}
\usepackage{nicefrac}   
\usepackage{cleveref}
\usepackage{natbib}
\usepackage{multirow}

\theoremstyle{plain}
\newtheorem{theorem}{Theorem}

\newtheorem{assumption}{Assumption}
\newtheorem{lemma}{Lemma}

\newtheorem{example}{Example}
\newtheorem{corollary}{Corollary}

\newcommand{\mG}{\mathcal{G}}
\newcommand{\mV}{\mathcal{V}}
\newcommand{\mE}{\mathcal{E}}
\newcommand{\mbR}{\mathbb{R}}

\newcommand{\mT}{\mathrm{T}}

\newcommand{\bB}{\bm{B}}
\newcommand{\bW}{\bm{W}}

\newcommand{\pkg}[1]{{\fontseries{b}\selectfont #1}}
\newcommand{\bY}{\bm{Y}}
\newcommand{\bX}{\bm{X}}
\newcommand{\bA}{\bm{A}}
\newcommand{\bmu}{\bm{\mu}}

\newcommand{\bC}{\bm{C}}

\newcommand{\bI}{\bm{I}}
\newcommand{\bE}{\bm{E}}
\newcommand{\bL}{\bm{L}}
\newcommand{\bSigma}{\bm{\Sigma}}
\newcommand{\bR}{\bm{R}}
\newcommand{\bOmega}{\bm{\Omega}}
\newcommand{\bdelta}{\bm{\delta}}

\newcommand{\bV}{\bm{V}}

\newcommand{\beginsupplement}{%
  \setcounter{figure}{0}%
  \setcounter{table}{0}%
  \setcounter{equation}{0}%
  \renewcommand{\thefigure}{S\arabic{figure}}%
  \renewcommand{\thetable}{S\arabic{table}}%
  \renewcommand{\theequation}{S\arabic{equation}}%
}

\title{Bayesian Causal Discovery with Cycles and Latent Confounders}

\author[1,$\dagger$]{Wei Jin}
\author[2,$\dagger$]{Lang Lang}
\author[3]{Amanda B. Spence}
\author[4,5]{Leah H. Rubin}
\author[2,*]{Yanxun Xu} 

\affil[1]{\small Department of Biostatistics, Boston University School of Public Health}
\affil[2]{\small Department of Applied Mathematics and Statistics, Johns Hopkins University}
\affil[3]{\small Department of Medicine, Georgetown University}
\affil[4]{\small Departments of Neurology, Psychiatry and Behavioral Sciences, Molecular and Comparative Pathobiology, Johns Hopkins University School of Medicine}
\affil[5]{\small Department of Epidemiology, Johns Hopkins Bloomberg School of Public Health}
\affil[$\dagger$]{\small \it These authors contributed equally to this work.}
\affil[*]{\small \it Correspondence should be addressed to email: yanxun.xu@jhu.edu}

\date{\vspace{-1cm}}

\begin{document}

\maketitle

\begin{abstract}
Learning causality from observational data has received increasing interest across various scientific fields. However, most existing methods assume the absence of latent confounders and restrict the underlying causal graph to be acyclic, assumptions that are often violated in many real-world applications. In this paper, we address these challenges by proposing a novel framework for causal discovery that accommodates both cycles and latent confounders. By leveraging the identifiability results from noisy independent component analysis and recent advances in factor analysis, we establish the unique causal identifiability under mild conditions. Building on this foundation, we further develop a fully Bayesian approach for causal structure learning, called BayCausal, and evaluate its identifiability, utility, and superior performance against state-of-the-art alternatives through extensive simulation studies. Application to a dataset from the Women’s Interagency HIV Study yields interpretable and clinically meaningful insights. To facilitate broader applications, we have implemented BayCausal in an R package, \pkg{BayCausal}, which is the first publicly available software capable of achieving unique causal identification in the presence of both cycles and latent confounders.
\end{abstract}


\section{Introduction}
Causal discovery, the task of inferring cause-effect relationships from data, is fundamental to scientific inquiry across disciplines.
With the rapid growth of real-world data sources, such as electronic health records, gene expression data, and health surveys, data-driven statistical and machine learning methods for causal discovery have become increasingly important across a broad range of scientific applications \citep{hill2016inferring,glymour2019review,addo2021exploring}. For example, in bioinformatics, reconstructing gene regulatory networks from gene expression data helps elucidate complex regulatory mechanisms within cellular systems \citep{zhang2012inferring}. In neuroscience, discovering causal relationships among biomarkers in patients with Alzheimer’s disease can provide valuable insights into the mechanisms driving disease progression \citep{shen2020challenges}. In atmospheric science, extracting causal relationships from complex, high-dimensional, and nonlinear earth system data can advance scientific knowledge of extreme weather events and broader climate dynamics \citep{runge2019inferring}.

Despite this growing demand, inferring causality from observational data remains challenging, particularly in the presence of latent confounders, i.e., unmeasured common causes of two or more observed variables. These confounders can introduce spurious associations that obscure the true underlying causal relationships. While numerous methods have been developed for discovering causality from observational data, including constraint-based \citep{spirtes2000causation}, score-based \citep{chickering2002optimal}, and functional-based approaches \citep{shimizu2006linear}, the vast majority rely on the causal sufficiency assumption, which presumes the absence of latent confounders. However, this assumption is hard to verify and is frequently violated in real-world applications. For example, in medical records, privacy concerns often restrict access to variables such as patients' socioeconomic status, which can confound both health outcomes and treatment assignments, leading to potential misinterpretations of treatment efficacy \citep{bernheim2008influence}. Failure to appropriately account for latent confounders can lead to biased, inaccurate, and misleading causal inferences. 

Another critical limitation of existing methods lies in the acyclicity assumption. Causal discovery methods typically aim to learn a directed graph, $\mG=(\mV,\mE)$, that represents the underlying causal relationships among a set of variables of interest. In the graph $\mG$, each directed edge points from the cause to the effect, e.g., for any vertices $X,Y\in \mV$, the edge $(X\to Y)\in \mE$ indicates that $X$ has a direct causal effect on $Y$ \citep{spirtes2000causation,pearl2009causality}. However, such graphical representations of causal mechanisms are often not uniquely identifiable from purely observational data. In general, they can be identified only up to a Markov equivalence class \citep{spirtes1995directed,koster1996markov,rantanen2020discovering,eberhardt2025discovering}, in which different graphs encode the same conditional independence relationships and thus cannot be distinguished from one another. To achieve unique identification, existing methods often rely on the assumption of acyclicity due to its convenient factorization properties, implying no directed cycles or feedback loops (e.g., $X \rightleftarrows Y$) in the directed graphs \citep{shimizu2006linear, hoyer2008nonlinear, peters2014identifiability}. While such methods offer theoretical and computational advantages, they fail to capture cyclic or reciprocal causal relationships, which are common in many real-world systems. For example, in the classic economic demand-supply model, demand directly influences supply, and supply, in turn, directly impacts demand, forming a feedback loop \citep{hyttinen2012learning}. Similarly, bidirectional causal relationships have been observed between insomnia and major psychiatric disorders \citep{gao2019bidirectional}.

Therefore, addressing both latent confounding and cyclic causal structures simultaneously is crucial for advancing causal discovery methods. While prior research has explored these challenges separately, including approaches that relax the acyclicity assumption \citep{lacerda2008discovering,mooij2011causal,mooij2020constraint,jin2025directed} and methods that accommodate latent confounders \citep{chen2013causality,cai2023causal,li2024nonlinear,chen2024discovery}, very few studies tackle them jointly. 
Theoretical results addressing these combined challenges exist in restricted cases. For example, \citet{hyttinen2012learning} established identifiability for linear cyclic models with latent confounders but relied on both observational and interventional data, the latter of which provides stronger causal evidence but may not always be practical or available. More recently, \citet{zhou2022causal} and \citet{li2024discovery} proposed methods to infer causal directions in bivariate graphs using heterogeneous observational data and invalid instrumental variables, respectively. While these approaches accommodate cycles and latent confounders between two variables, it remains unclear how to extend them to handle multivariate graphs.

In this paper, we propose a novel framework for causal discovery that simultaneously addresses latent confounders and cyclic causal structures using purely observational data. Our key identification strategy relies on the distributional assumptions that the primary observed variables of interest are non-Gaussian, while the latent confounders follow a Gaussian distribution. By leveraging the identifiability results from noisy independent component analysis (ICA; \citealt{comon1994independent,davies2004identifiability}) under these distributional assumptions, we prove that the directed graph representing the causal relationships among the primary observed variables is uniquely identifiable. Additionally, with mild assumptions on error structures, we extend identifiability to the latent confounders themselves by leveraging recent advances in factor analysis \citep{fruhwirth2024sparse}. Furthermore, we develop a fully Bayesian approach for causal structure learning, called BayCausal, which not only infers the underlying causal graph with natural uncertainty quantification but also enables data-driven inference of the unknown number of latent confounders through a reversible jump Markov Chain Monte Carlo (RJMCMC) algorithm. For reproducibility and broader dissemination, we build an R package, \pkg{BayCausal}, that implements BayCausal available at \url{https://github.com/thepianistalex/BayCausal}. To the best of our knowledge, \pkg{BayCausal} is the first publicly available software for causal discovery that achieves unique identifiability in the presence of both cycles and latent confounders. Through simulations and an application to a real-world HIV database, i.e., the Women's Interagency HIV Study (WIHS, \citealt{adimora2018cohort}), we demonstrate the identifiability, utility, and superiority of BayCausal against state-of-the-art alternatives.

The remainder of this paper is organized as follows. In Section~\ref{sec:model}, we describe the data-generating model of the proposed framework, which accounts for both cycles and latent confounders. Section~\ref{sec:theory} establishes the unique causal identifiability of the proposed model. In Section~\ref{sec:mcmc}, we present the Bayesian structural learning approach, BayCausal, for estimating the proposed model. Section~\ref{sec:simu} empirically validates the proposed causal identification theory and compares the performance of BayCausal with state-of-the-art alternatives through extensive simulation studies. In Section~\ref{sec:wihs}, we demonstrate the practical utility of BayCausal by applying it to the WIHS dataset. Lastly, we conclude with a discussion in Section~\ref{sec:con}.

\section{Data-Generating Model}
\label{sec:model}
Suppose there are $n$ independent and identically distributed observations from a population. For the $i$-th observation, let $\bY_{i}\in \mbR^Q$ denote the primary variables of interest, whose underlying causal relationships we aim to uncover. Let $\bX_i$ denote an $S$-dimensional vector representing the covariates (e.g., age and race) for the primary variables, which can be either continuous or discrete. We denote the latent confounders as $\bC_i \in \mathbb{R}^P$. 
The causal directions between the covariates and the primary variables, as well as those between the latent confounders and the primary variables, are both fixed to be from the former to the latter \textit{a priori}. For notational simplicity, the observation index $i$ will be suppressed when understood from the context. 

To investigate the underlying causal relationships among the primary variables while accounting for the effects of covariates and latent confounders, we propose the following data-generating model:
\begin{equation}
    \label{eqn:full_model}
    \bY = \bmu + \bB\bY + \bA\bX + \bL\bC + \bE,
\end{equation}
where $\bB$ is a $Q \times Q$ matrix of causal effects among the primary variables, $\bA$ is a $Q \times S$ matrix of covariate effects, $\bL$ is a $Q \times P$ matrix of latent confounding effects, where $P < Q$. The assumption of $P < Q$ is common in causal discovery, where latent confounders represent a reduced set of hidden factors that jointly influence the observed primary variables \citep{chen2024discovery}. It ensures the model's parsimony and identifiability while capturing the shared variability among the observed variables. 
Additionally, $\bmu\in \mbR^{Q}$ is the intercept, and $\bE \in\mbR^{Q}$ is the exogenous error. We further assume that the latent confounders $\bC$ are independent of both covariates and exogenous errors, and they have zero mean and unit variance. By standardizing the latent confounders in this way, their average effects and magnitude are explicitly captured by the global intercept $\bmu$ and the loading matrix $\bL$, respectively. 

Equation~\eqref{eqn:full_model} represents a structural causal model (SCM; \citealt{bollen1989structural,pearl2009causality}) that describes how the primary variables $\bY$ are generated by their direct causes. Specifically, $Y_{q'}$ is a direct cause of $Y_q$ if and only if $B_{qq'}\neq 0$. Likewise, $X_s$ and $C_p$ are direct causes of $Y_q$ if and only if $A_{qs}\neq 0$ and $L_{qp}\neq 0$, respectively. Note that Equation~\eqref{eqn:full_model} can also be expressed as a fixed-point equation in the form of $\bY = \bB\bY + \bm{D}$, where $\bm{D}$ includes the intercept $\bmu$, the covariate effects $\bA\bX$, the confounding effects $\bL\bC$, and the exogenous error $\bE$. To ensure that the proposed model is well-defined, i.e., there exists a unique solution $\bY^{*}$ that satisfies the fixed-point equation $\bY^{*} = \bB\bY^{*} + \bm{D}$, we impose the following \textit{stability} condition on $\bB$. Specifically, the matrix $\bB$ is stable if the maximum modulus of its eigenvalues is strictly less than $1$, indicating that there exists a matrix norm $\|\cdot\|$ such that $\|\bB\| < 1$. Consequently, the above fixed-point equation is a contraction. By the Banach fixed-point theorem \citep{berinde2007iterative}, a unique fixed-point solution is guaranteed. 

\begin{figure}[!htb]
  \centering
  \includegraphics[width=\textwidth]{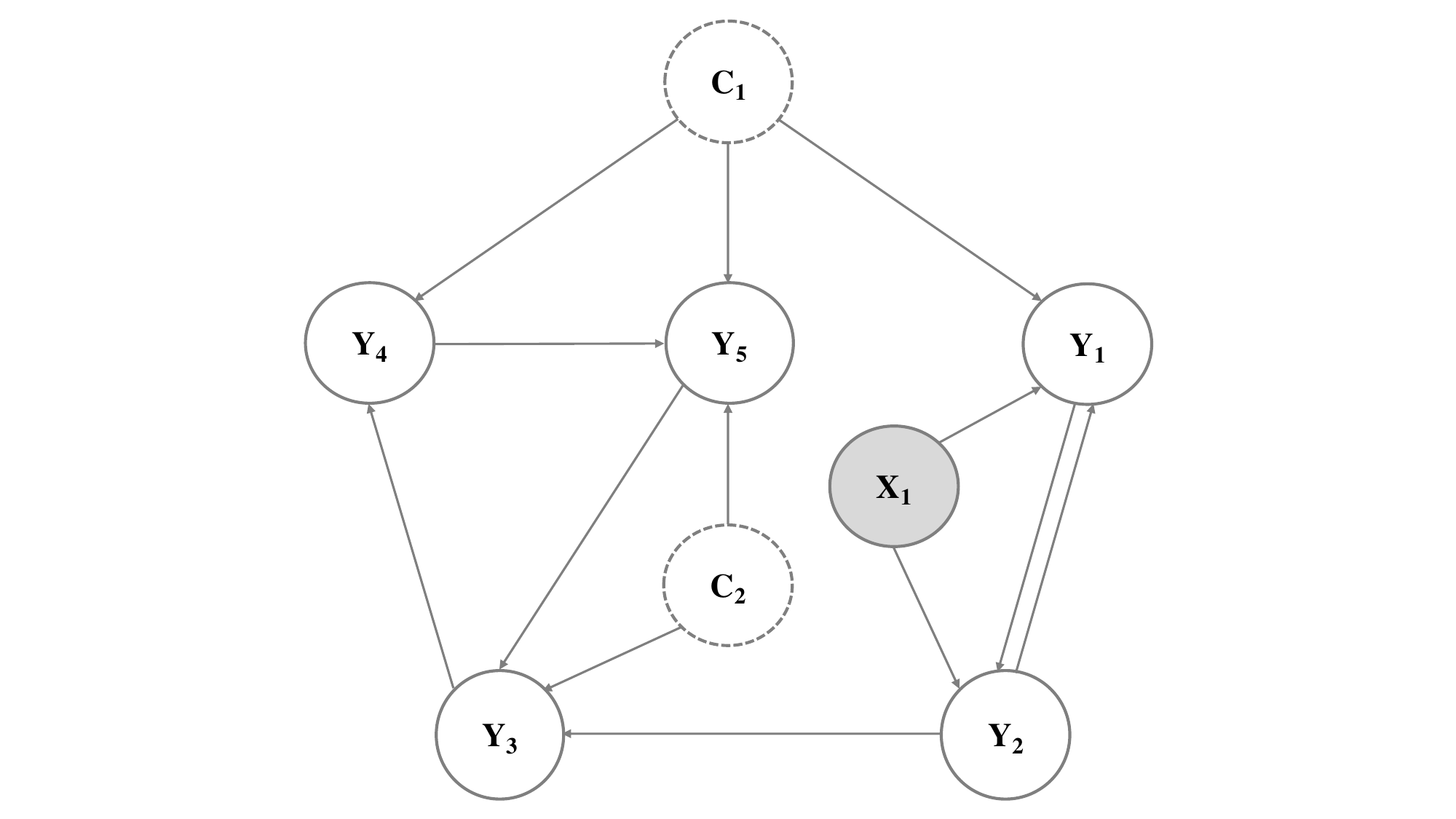}
  \caption{A graphical illustration of the proposed data-generating model. The white solid circles represent the primary variables, the gray solid circles represent the covariates, and the white dashed circles represent the latent confounders.}
  \label{fig:graph_example}
\end{figure}

The proposed model can also be represented as a directed graph $\mG = (\mV,\mE)$, where $\mV = \{\bY, \bX, \bC\}$ denotes the set of vertices representing the variables of interest, and $\mE \subset \mV \times \mV$ denotes the set of directed edges capturing the direct causal relationships between these variables. A graphical illustration of the proposed model is provided in Figure~\ref{fig:graph_example}. For any two vertices $V_{k}, V_{k'}\in \mV$, if $(V_{k} \to V_{k'})\in\mE$, then there exist a direct causal relationships from the parent $V_{k}$ to the child $V_{k'}$. For any $K\geq 2$ vertices $V_{1},\dots, V_{K}\in \mV$, if $(V_{1} \to V_{2})\in\mE$, $(V_{2} \to V_{3})\in\mE$, $\dots$, and $(V_{K-1} \to V_{K})\in\mE$, then there exits a directed path from the ancestor $V_{1}$ to the descendant $V_{K}$. A special case occurs when $V_1=V_K$, where the path $V_{1} \to V_{2}\to \dots \to V_{K-1} \to V_1$ forms a \textit{cycle}. Two cycles are \textit{disjoint} if they do not share any common vertices. By definition, cycles are allowed among primary variables in the proposed model. For example, there exist two disjoint cycles $Y_1 \rightleftarrows Y_2$ and $Y_3\to Y_4\to Y_5 \to Y_3$ in Figure~\ref{fig:graph_example}, indicating that $B_{21}, B_{12}\neq 0$ and $B_{43}, B_{54}, B_{35}\neq 0$ in \eqref{eqn:full_model}. However, \textit{self-loops} (i.e., $(Y_q \to Y_q)\in \mE$ or $B_{qq}\neq 0$ for any $q$) are not allowed by convention.

To model latent confounders, we adopt the \textit{canonical} form commonly used in the causal discovery literature \citep{hoyer2008estimation,cai2023causal}. Specifically, we assume that latent confounders can only be the parents of primary variables. In addition, each latent confounder has no parents but must have at least two children, and no latent confounders share exactly the same sets of children. 
The canonical form ensures that the latent variables $\bC$ serve as valid confounders of primary variables $\bY$, as they act as parents (i.e., common causes) for at least two primary variables in the causal graph. It also avoids redundancy in the latent variable modeling by preventing latent confounders from having parents or identical sets of children, thereby enhancing the ability to disentangle the influence of latent confounders on primary variables. For example, the graphical model with two latent confounders $C_1$ and $C_2$ shown in Figure~\ref{fig:graph_example} follows the canonical form. More graphical examples following the canonical and non-canonical forms can be found in Supplementary Figure S1.

\section{Causal Identification Theory}
\label{sec:theory}
In this section, leveraging identifiability results from noisy ICA and recent advances in factor analysis (FA), we establish the first causal identification theory for directed graphs with both cycles and latent confounders that achieves unique identifiability from observational data. We begin with a brief overview of our strategy that leads to causal identification of the proposed data-generating model, followed by detailed explanations. 

\subsection{Overview}
The fundamental challenge in causal discovery is to determine under which assumption one can uniquely identify a graphical representation of the underlying causal mechanisms from observational data. Despite extensive research on causal identifiability for directed graphs with either cycles \citep{lacerda2008discovering,roy2023directed,jin2025directed} or latent confounding \citep{chen2013causality,salehkaleybar2020learning,wang2023causal}, no existing results achieve unique causal identifiability when both challenges are present. This is a critical gap that we aim to address in this work. 

To establish the causal identification theory, we begin by considering a simplified version of the proposed data-generating model \eqref{eqn:full_model}, where intercepts and covariates that do not affect identifiability are omitted: 
\begin{equation}
    \label{eqn:simple_model}
    \bY = \bB\bY + \bL\bC + \bE. 
\end{equation}
Our main strategy is to decompose the complex task of establishing causal identifiability for model \eqref{eqn:simple_model} into two key sub-tasks. First, we identify the causal relationships among the primary variables encoded by $\bB$. Leveraging identifiability results from noisy ICA, we show that $\bB$ is uniquely determined when the graph contains only disjoint cycles and the exogenous errors are non-Gaussian. Second, once $\bB$ is identified, we recast \eqref{eqn:simple_model} in an equivalent FA form as $\widetilde{\bY} = \bL\bC + \bE$, where $\widetilde{\bY} = (\bI - \bB)\bY$ are the transformed variables. This representation allows us to recover the latent confounder effects $\bL$. Finally, we extend these results to the full model by reintroducing intercepts and covariates.

\subsection{\mbox{Causal identification among primary variables}}
Without additional assumptions, directed cyclic graphs are only identifiable up to Markov equivalence classes \citep{spirtes1995directed,koster1996markov}.   To achieve unique causal identifiability, we impose the following standard assumptions, widely used in causal discovery.

\begin{assumption}
    \label{asm:causal}
    \hfill
    \begin{itemize}
        \item[(a)] Gaussian confounders: The latent confounders $\bC\in \mathbb{R}^{P}$ follow a multivariate Gaussian distribution $\mathcal{N}(\bm{0},\bI_{P})$.
        \item[(b)] Non-Gaussian errors: The jointly independent exogenous errors $\bE \in \mathbb{R}^{Q}$ are all non-Gaussian distributed.
    \end{itemize}
\end{assumption}

The assumption of Gaussian confounders has been widely adopted in the existing literature \citep{chen2013causality,chen2024discovery,li2024nonlinear}. This choice is motivated by the fact that $\bC$ aggregates many unmeasured factors and, by the central limit theorem, is well approximated by a Gaussian distribution. The non-Gaussian-error assumption is likewise standard \citep{shimizu2006linear,salehkaleybar2020learning,jin2025directed}, motivated by the prevalence of non-Gaussian variables in practice and their favorable identifiability properties: non-Gaussianity provides higher-order statistical information (e.g., nonzero cumulants) that breaks the symmetry of Gaussian noise, enabling unique causal identification.
 
Under Assumption~\ref{asm:causal}, model~\eqref{eqn:simple_model} can be rewritten in a noisy ICA form
$\bY = \widetilde{\bB}(\bL\bC + \bE)$, where $\widetilde{\bB} = (\bI - \bB)^{-1}$ is the \textit{mixing} matrix. The goal of noisy ICA is to uniquely identify a mixing matrix that maps the latent non-Gaussian sources $\bE$ to the observed signals $\bY$, while allowing for the presence of additional Gaussian noises through $\bC$. The noisy ICA framework ensures that the mixing matrix $\widetilde{\bB}$ is identifiable up to permutation and scaling. We further show that for any directed graph that does not contain joint cycles, there exists a unique stable solution for $\bB$ among all possible permutations and scalings of the mixing matrix. This leads to our causal identifiability result for the causal relationships among primary variables, summarized in the following lemma.

\begin{lemma}[Identification of the causal relationships among primary variables]
    \label{lem:identification_B}
    Consider model~\eqref{eqn:simple_model} and suppose that Assumptions~\ref{asm:causal}(a)-(b) hold. For any directed graph $\mG$ associated with model~\eqref{eqn:simple_model} that contains only disjoint cycles, there is a unique stable $\bB$ consistent with $\mG$.
\end{lemma}

Lemma~\ref{lem:identification_B} implies that, among all possible directed graphs that contain only disjoint cycles, the corresponding matrix $\bB$, which represents the causal relationships among the primary variables, can be uniquely identified from observational data. We leave the detailed proof of Lemma~\ref{lem:identification_B} to the Supplementary Materials Section A1. 

\subsection{Identification of the latent confounding effects}
With $\bB$ identified, we rewrite model~\eqref{eqn:simple_model} as
 $\widetilde{\bY} = \bL\bC + \bE$, where $\widetilde{\bY} = \widetilde{\bB}^{-1}\bY$ contains all the identified components. 
The goal of FA is to determine the dependence structure among the identified observations $\widetilde{\bY}$ through the decomposition of the covariance matrix of $\widetilde{\bY}$ as $\bOmega = \bL\bL^{\mT} + \bSigma_{E}$, where $\bL$ is the factor loading matrix and $\bSigma_{E}$ is a diagonal matrix with nonnegative diagonal entries. In this section, we first introduce two major non-identifiability issues associated with FA, and then present our strategies to resolve these challenges in order to achieve unique identification of the latent confounding effects $\bL$.

The first non-identifiability issue is the indeterminacy between the diagonal elements of $\bL\bL^{\mT}$ and $\bSigma_{E}$ \citep{davies2004identifiability}. Since all off-diagonal entries of $\bSigma_{E}$ are zero, the off-diagonal entries of $\bL\bL^{\mT}$ can be uniquely determined by those of the observed covariance matrix $\bOmega$. However, the diagonal elements (i.e., the variance terms) of $\bL\bL^{\mT}$ and $\bSigma_{E}$ are not uniquely identifiable in general. Consider the following simple example for illustration.

\begin{example}
    Consider model~\eqref{eqn:simple_model} in the following FA form:
    \begin{equation*}
        \widetilde{\bY} = \bL\bC + \bE = \begin{bmatrix} 2 & 0 \\ 3 & 1\end{bmatrix} \begin{bmatrix} C_1 \\ C_2 \end{bmatrix} + \begin{bmatrix} E_1 \\ E_2 \end{bmatrix}.
    \end{equation*}
    This model can be re-expressed in an equivalent form by incorporating one of the independent Gaussian components, $C_2$, into the independent noise term $E_2$ as follows:
     \begin{equation*}
        \widetilde{\bY} = \bL'\bC' + \bE' = \begin{bmatrix} 2 \\ 3\end{bmatrix} C_1  + \begin{bmatrix} E_1  \\ C_2+E_2 \end{bmatrix}.
    \end{equation*}
     Therefore, the diagonal elements of $\bL\bL^{\mT}$ and $\bSigma_{E}$ are not uniquely identifiable, as different allocations of variance between the factor loading matrix and the exogenous error term can yield the same observed covariance.
\end{example}

In the FA literature, a common strategy to resolve this non-identifiability issue is to impose the so-called \textit{3579 counting rule} \citep{anderson1956statistical,sato1992study,fruhwirth2023counts} on the factor loading matrix $\bL$. Specifically, a $(Q \times P)$-dimensional factor loading matrix $\bL$ satisfies the 3579 counting rule if, for each submatrix consisting of $p$ columns of $\bL$, the number of nonzero rows in the submatrix is at least $2p+1$, for $p=1,2,\dots,P$. In other words, the 3579 counting rule requires that every column of $\bL$ contains at least 3, every pair of columns shares at least 5, every subset of 3 columns shares at least 7 nonzero entries, and so forth. The 3579 counting rule essentially needs sufficient information from the off-diagonal elements of the matrix $\bL\bL^{\mT}$ to constrain the degrees of freedom in its diagonal elements, thereby leading to the unique identifiability. However, this condition implicitly requires that the matrix $\bL$ is not too sparse, which is reasonable in the context of FA but may not be appropriate for our setting. In the context of FA, the correlations among observed variables are fully explained by latent confounders. In contrast, in causal discovery, correlations among observed variables are typically attributed to their direct causal relationships; therefore, it is often unrealistic to assume that a large number of observed variables are always confounded by certain latent factors. 

The second non-identifiability issue is the rotational indeterminacy of $\bL$ \citep{geweke1996measuring}. In particular, for any orthogonal matrix $\bR$ such that $\bR\bR^{\mT}=\bI$, the transformed loading matrix $\widetilde{\bL}=\bL\bR$ results in the same representation of the matrix $\bL\bL^{\mT}$ as $\bL$. The most popular solution to address the rotational indeterminacy is to require $\bL$ to follow a lower triangular structure with positive diagonal elements (PLT; \citealt{joreskog1969general,carvalho2008high,bhattacharya2011sparse}). However, the PLT structure is quite restrictive, as it induces order dependence among the responses, making the selection of the first few responses an important modeling decision. Furthermore, the PLT structure is not general enough, in the sense that not every factor loading matrix can be rotated into a PLT structure.

To address these two major challenges in uniquely identifying the matrix $\bL$, we introduce the following assumptions.

\begin{assumption}
    \label{asm:factor_analysis}
    \hfill
    \begin{itemize}
        \item[(a)] Distributional assumption: The non-Gaussian distribution of the exogenous error $E_q$, $q=1,2,\dots,Q$, is not closed under Gaussian convolutions. Specifically, for any Gaussian random variable $C_p$ independent of $E_q$, $p=1,2,\dots,P$, $C_p+E_q$ does not follow the same non-Gaussian distribution form as $E_q$. 
        
        \item[(b)] Unordered generalized lower triangular (UGLT) structure: For each column $p$ of the $(Q \times P)$-dimensional factor loading matrix $\bL$ with full column rank, $p=1,2,\dots,P$, the row index of the first non-zero entry in column $p$ is called the pivot row, denoted as $\ell_p$ (i.e., $L_{qp} = 0$ for all $q < \ell_p$ and $L_{\ell_p,p} \neq 0$). Then the factor loading matrix $\bL$ has an unordered generalized lower triangular (UGLT) structure if the pivot rows $\ell_1,\dots,\ell_P$ are pairwise distinct, i.e., $\ell_p \neq \ell_{p'}$ for any $p\neq p'$.
    \end{itemize}
\end{assumption}

The indeterminacy between the diagonal elements of $\bL\bL^{\mT}$ and $\bSigma_{E}$ can be addressed under Assumption~\ref{asm:factor_analysis}(a). This is because, the non-Gaussian distribution of $\bE$ is not closed under Gaussian convolution, leaving no flexibility to reallocate any Gaussian components from the latent confounders to the exogenous errors. As a result, the variance terms of both $\bL\bL^{\mT}$ and $\bSigma_{E}$ are uniquely identifiable. Assumption~\ref{asm:factor_analysis}(a) is a mild distributional condition satisfied by many common non-Gaussian distributions. The following simple example illustrates that the Laplace distribution satisfies this assumption.

\begin{example}
    Consider a univariate zero-mean Laplace random variable $E$ with scale parameter $b>0$, and an independent univariate zero-mean Gaussian random variable $C$ with variance $\sigma^2>0$. Then the characteristic function of $Z=E+C$ is given by $\phi_Z(t)=\phi_E(t)\phi_C(t)=\frac{e^{-\sigma^2 t^2/2}}{1 + b^2t^2}$. Suppose that $Z$ is also Laplace distributed with some scale parameter $b'>0$, then its characteristic function would take the form $\phi_Z(t)=\frac{1}{1 + b'^2t^2}$. For both expressions of $\phi_Z(t)$ to be identical for all $t$, it must hold that $e^{-\sigma^2 t^2/2}=\frac{1 + b^2 t^2}{1 + b'^2 t^2}$. However, the left-hand side is an exponential function, whereas the right-hand side is a rational function. These two forms can only be equal for all $t$ if $\sigma^2=0$, which is a contradiction. 
\end{example}

In fact, it can be easily shown that common non-Gaussian distributions such as the Uniform, Cauchy, Logistic, and Student's $t$-distributions also satisfy Assumption~\ref{asm:factor_analysis}(a). Detailed proofs are provided in the Supplementary Materials Section A2.

The UGLT structural constraint on the factor loading matrix (i.e., Assumption~\ref{asm:factor_analysis}(b)) has been recently introduced in the FA literature \citep{fruhwirth2024sparse}, which strikes a balance between ensuring identifiability of the factor loading matrix and keeping structural constraints as weak as possible. On the one hand, the UGLT structure is a much weaker condition than the commonly adopted PLT structure, allowing any factor loading matrix to be rotated into a UGLT representation. The unordered nature of UGLT also helps mitigate the order dependencies among the responses introduced by the PLT constraint. On the other hand, it is strong enough to resolve the non-identifiability issue arising from the rotational indeterminacy. The following example illustrates the difference between the UGLT structure and the PLT structure constraints on the factor loading matrix.

\begin{example}
Consider the following factor loading matrices with $Q=5$ response variables and $P=2$ latent factors:
\begin{equation*}
\bL_{UGLT}=
\begin{pmatrix}
L_{11} & 0 \\
L_{21} & 0 \\
L_{31} & 0 \\
0 & L_{42} \\
0 & L_{52} 
\end{pmatrix},
\quad
\bL_{PLT}=
\begin{pmatrix}
L_{11} & 0 \\
L_{21} & L_{22} \\
L_{31} & 0 \\
0 & L_{42} \\
0 & L_{52} 
\end{pmatrix}, 
\end{equation*}
where $L_{qp}$ indicates non-zero entries. The factor loading matrix $\bL_{UGLT}$ has a UGLT structure, and $\bL_{PLT}$ has a PLT structure. Note that $\bL_{UGLT}$ is not a PLT matrix since $L_{22}=0$, whereas $\bL_{PLT}$ is a UGLT matrix because its pivot rows satisfy $\ell_1 =1 \neq \ell_2=2$. 

To give an intuitive demonstration that UGLT is a weaker condition than PLT, we show that $\bL_{UGLT}$ can not be rotated into a PLT matrix form. Specifically, for any rotation matrix
\begin{equation*}
\bR_{\alpha \beta} = 
\begin{pmatrix}
\cos\alpha  & (-1)^{\beta}\sin\alpha \\[2pt]
-\sin\alpha & (-1)^{\beta}\cos\alpha
\end{pmatrix},
\quad\alpha\in[0,2\pi),\; \beta\in\{0,1\},
\end{equation*}
we apply it to $\bL_{UGLT}$ and obtain the rotated matrix
\begin{equation*}
\bL_{UGLT}\bR_{\alpha\beta}
=
\begin{pmatrix}
 L_{11}\cos\alpha  & (-1)^{\beta}L_{11}\sin\alpha \\[2pt]
 L_{21}\cos\alpha  & (-1)^{\beta}L_{21}\sin\alpha \\[2pt]
 L_{31}\cos\alpha  & (-1)^{\beta}L_{31}\sin\alpha \\[2pt]
-\!L_{42}\sin\alpha & (-1)^{\beta}L_{42}\cos\alpha \\[2pt]
-\!L_{52}\sin\alpha & (-1)^{\beta}L_{52}\cos\alpha
\end{pmatrix}.
\end{equation*}
Suppose $\bL_{UGLT}\bR_{\alpha\beta}$ is in PLT form, then the above-diagonal element $(-1)^{\beta}L_{11}\sin\alpha$ has to be zero, which implies that $\sin\alpha =  0$. However, this will also force the second diagonal element to be zero, leading to a contradiction.
\end{example}

Under these two assumptions (i.e., Assumptions~\ref{asm:factor_analysis}(a)-(b)), the latent confounding effects $\bL$ can be identifiable up to column permutations and sign flips. Without loss of generality, we conclude that $\bL$ is uniquely identified. This is because column permutations and sign flips correspond to label switching and sign indeterminacy of the latent confounders $\bC$, and the ordering and sign of the columns carry no meaningful interpretation as they are latent. We summarize the identification result for the latent confounding effects $\bL$ in the following lemma, the detailed proof of which is provided in the Supplementary Materials Section A3.

\begin{lemma}[Identification of the latent confounding effects]
    \label{lem:identification_L}
    Consider model~\eqref{eqn:simple_model} and suppose that Assumptions~\ref{asm:causal}(a)-(b) and Assumptions~\ref{asm:factor_analysis}(a)-(b) hold, then the latent confounding effects $\bL$ are uniquely identifiable from observational data.
\end{lemma}

By combining the identification results for the causal relationships among primary variables and the latent confounding effects, we establish the causal identification theory for 
model~\eqref{eqn:simple_model}, as stated in the following theorem.

\begin{theorem}[Causal Identification]
    \label{thm:causal}
     Suppose Assumptions~\ref{asm:causal}(a)–(b) and Assumptions~\ref{asm:factor_analysis}(a)–(b) hold, then there exists a unique directed graph with disjoint cycles consistent with model~\eqref{eqn:simple_model}. In addition, the causal relationships among primary variables $\bB$ and the latent confounding effects $\bL$ are uniquely identifiable from observational data.
\end{theorem}

Theorem~\ref{thm:causal} follows directly from Lemma~\ref{lem:identification_B} and Lemma~\ref{lem:identification_L}, and thus its proof is omitted.

\subsection{Causal identification for full model}
We now establish the identifiability of the proposed data-generating model~\eqref{eqn:full_model} in the presence of both cycles and latent confounders by adding back the intercepts and covariates to the simplified model~\eqref{eqn:simple_model}.  

\begin{corollary}[Causal Identification]
    \label{cor:causal}
    Suppose Assumptions~\ref{asm:causal}(a)–(b) and Assumptions~\ref{asm:factor_analysis}(a)–(b) hold, then there exists a unique directed graph with disjoint cycles consistent with the proposed data-generating model~\eqref{eqn:full_model}. In addition, the parameters of model~\eqref{eqn:full_model} are also uniquely identifiable from observational data.
\end{corollary}

In the proof of Corollary~\ref{cor:causal}, we demonstrate that the inclusion of intercepts and covariates does not affect the identifiability results established in Theorem~\ref{thm:causal}. We provide the detailed proof of Corollary~\ref{cor:causal} in the Supplementary Materials Section A4.
 
\section{Bayesian Structural Learning}
\label{sec:mcmc}
In this section, we propose a Bayesian structural learning approach, called BayCausal, to estimate the causal graph associated with the proposed data-generating model~\eqref{eqn:full_model}, using posterior inference via a Markov chain Monte Carlo (MCMC) algorithm. We begin by introducing prior specifications that impose sparsity structures on both causal relationships among primary variables $\bB$ and latent confounding effects $\bL$ to enhance interpretability of the inferred causal graph. We then present a reversible jump MCMC (RJMCMC) algorithm that automatically infers the number of latent confounders while quantifying associated uncertainty, an important quantity that is typically unknown in practice.

\subsection{Imposing sparsity on the causal graph}
To encourage sparsity in the inferred causal graphs, we consider spike-and-slab priors \citep{ishwaran2005spike}. 
For the causal effect of primary variable $Y_{q'}$ on $Y_q$, we assume $B_{qq'} \sim \mathcal{N}(0, \gamma_{qq'}\nu_{qq'})$, where $\nu_{qq'}\sim \text{Inverse-Gamma}(a_{v},b_{v})$ and $\gamma_{qq'} \sim \rho \Delta_{1}(\gamma_{qq'})+ (1 - \rho) \Delta_{\nu_0}(\gamma_{qq'})$. Here $\Delta_x(\cdot)$ denotes the Dirac measure at $x$, and $\nu_{0}$ is a small pre-specified constant (e.g., $\nu_{0}=2.5\times 10^{-4}$). When $\gamma_{qq'} = 1$ (i.e., slab),  $B_{qq'}$ is nonzero, indicating that $Y_{q'}$ has a causal effect on $Y_q$; when $\gamma_{qq'} = \nu_0$ (i.e., spike), $B_{qq'}$ is negligible, indicating no significant causal effect. We further assume that $\rho \sim \text{Beta}(a_\rho,b_\rho)$, following \cite{scott2010bayes}. We assign analogous spike-and-slab priors to covariate effects $A_{qs}$, for $q=1,2,\dots,Q$ and $s=1,2,\dots,S$.

For the latent confounding effects $\bL$, let $P^*$ denote the underlying true number of latent confounders that we aim to identify from data. For now, we treat $P^*$ as fixed; the RJMCMC algorithm for inferring $P^*$ from data will be described later. Note that in the proposed model~\eqref{eqn:full_model}, $P$ indicates the maximum number of latent confounders to be considered, i.e., $0\leq P^* \leq P < Q$. In other words, the first $P^*$ columns of $\bL$ are non-zero, while the remaining $P-P^*$ columns are zero columns. We assign the following spike-and-slab priors, $L_{qp} \mid \zeta_p \sim (1 - \zeta_p)\Delta_{0} + \zeta_p \mathcal{N}(0, \kappa \sigma_q^{2})$, where $\sigma_q^{2}$ denotes the variance term for the exogenous error $E_q$. Here $\kappa$ is a global shrinkage parameter that follows $\text{Inverse-Gamma}(a_\kappa, b_\kappa)$. Following  \cite{fruhwirth2023generalized}, we assign a finite two-parameter-beta prior on $\zeta_{p}$, i.e., $\zeta_{p} \sim \text{Beta}\Bigl(\frac{a_{1}a_{2}}{P},a_{2}\Bigr)$, for $p=1,2,\dots,P$, where the hyperparameters $a_{1}, a_{2}$ are assumed to follow $\text{Inverse-Gamma}(b_{1},c_{1})$ and $\text{Inverse-Gamma}(b_{2},c_{2})$, respectively.

Beyond sparsity, two structural assumptions are essential for the identifiability of the latent confounding effects $\bL$ (i.e., Assumptions~\ref{asm:factor_analysis}(a)-(b)): non-Gaussian errors and the UGLT structure. For the non-Gaussian error, we assume a Laplace distribution: $E_{q} \sim \text{Laplace}(0, 2\sigma_q)$, $q=1,2,\dots,Q$. This choice simplifies posterior computation as the Laplace distribution can be represented as a continuous scale mixture of normal distributions. Specifically, by introducing the auxiliary variable $\tau_{q} \sim \text{Inverse-Gamma}(1, 1/8)$ and assuming $E_{q}\mid\tau_{q}\sim \mathcal{N}\bigl(0,\sigma_q^2/\tau_{q}\bigr)$, the marginal distribution $E_{q}$ is Laplace \citep{choi2013analysis}. We complete this specification with an Inverse-Gamma prior on the error variance: $\sigma_q^2 \sim \text{Inverse-Gamma}(a_\sigma, b_\sigma)$.

To enforce the UGLT structure, we specify the priors for the $P^*$ pivot rows $\bm{\ell}=(\ell_1,\ell_2,\dots,\ell_{P^*})$, one for each non-zero column of $\bL$. Note that these pivot rows must be pairwise distinct. We assume a uniform prior for the selection of each pivot row $\ell_{p}$, given by $p\bigl(\ell_{p} \mid \bm{\ell}_{-p}\bigr)=1/\lvert \mathcal{Q}(\bm{\ell}_{-p}) \rvert=1/(Q-P^*+1)$, where $\mathcal{Q}(\bm{\ell}_{-p})$ is the set of unused row indices, thereby ensuring each available row is equally likely to be selected as a pivot for the current factor column $p$. Subsequently, we define the $(Q\times P^*)$-dimensional sparsity matrix $\bdelta$, where $\delta_{qp}=1$ if $L_{qp} \neq 0$. Conditional on the pivot row $\ell_p$ for column $p$, the prior for $\mathbf{\delta}$ enforces the UGLT constraints: entries above the pivot must be zero (i.e., $P(\delta_{qp}=1 \mid \ell_p)=0$ for $q<\ell_p$), the pivot entry itself must be non-zero (i.e., $P(\delta_{qp}=1 \mid \ell_p)=1$ for $q=\ell_p$), and entries below the pivot are stochastically included with probability $\zeta_p$ (i.e., $P(\delta_{qp}=1 \mid \ell_p)=\zeta_p$ for $q>\ell_p$). 

Lastly, we complete the prior specification by assigning $\mu_q \sim \mathcal{N}(0,\sigma_{\mu}^2)$ with a closed-form solution for ease of posterior computation. 

\subsection{Searching for the number of latent confounders}
We have so far assumed a fixed number of latent confounders $P^*$. However, in real-world applications, $P^*$ is typically unknown \textit{a priori}. To accurately capture the structure of latent confounders in the causal graph, it is essential to determine $P^*$ in a data-driven manner. By leveraging the Bayesian framework, we introduce an RJMCMC algorithm that efficiently explores the model space across varying numbers of latent confounders with natural uncertainty quantification. The RJMCMC algorithm modifies the number of latent confounders through two core moves: split and merge. Specifically, in a split move, a zero column of $\bL$ is converted into a non-zero column with a certain probability. Conversely, in a merge move, a non-zero column of $\bL$ is converted into a zero column with a certain probability.

With a slight abuse of notation, we again use $P^*$ to denote the number of latent confounders identified by the RJMCMC algorithm at the current stage. Define $P_{single}$ as the number of columns with only one non-zero entry in the sparsity matrix $\bdelta$. Then conditional on $P^*$ and $P_{single}$, let $p_{split}(P^*,P_{single})$ and $p_{merge}(P^*,P_{single})$ denote the probabilities of splitting and merging at the current stage, respectively. These probabilities are naturally constrained.  
In particular, when $P^* = P$, the maximum number of latent confounders has been reached, so no split move is considered, and we set $p_{split}(P^*,P_{single}) = 0$. When $P_{single} = 0$ or $P^* = P_{single} = 1$, the merge move is not considered and we set $p_{merge}(P^*,P_{single}) = 0$. When both moves are possible, we employ a symmetric proposal distribution with equal probability for splitting and merging.  Specifically, a split move turns one of the $P-P^*$ zero columns in $\bdelta$ into a column with only one non-zero entry, with proposal density $q_{split}\bigl(\bdelta^{new}\mid \bdelta\bigr) = \frac{1}{2(P - P^*)}$. A merge move turns one of the $P_{single} > 0$ columns with only one non-zero entry in $\bdelta$ into a zero column, with proposal density $q_{merge}\bigl(\bdelta^{new}\mid \bdelta\bigr) = \frac{1}{2P_{single}}$. 

When proposing a split move on one of the $P-P^*$ zero columns $p$ in $\bdelta$, we first sample a pivot row $\ell_p$ for column $p$ from the set of pivot positions not yet assigned to the existing $P^*$ non-zero columns. We then define the following transformation between different model spaces, mapping from $(\sigma_{\ell_p}^2, U)$ to $((\sigma_{\ell_p}^2)^{new}, L_{\ell_{p},p}^{new})$, specified as $(\sigma_{\ell_p}^2)^{new} = (1 - U^2)\sigma_{\ell_p}^2$ and $L_{\ell_{p},p}^{new} = \sqrt{8\sigma_{\ell_p}^2}U$, where $U$ is an auxiliary variable sampled from the uniform distribution on the interval $(0,1)$. After this transformation, the diagonal entries of $\bL\bL^{\mT}+\bSigma_E$ remain unchanged under the assumption of Laplace-distributed errors $\bE$, which facilitates a higher acceptance rate by preserving the marginal variances of the data across different models. The acceptance probability for the split move is given by $\min(1,MH_{split}(P^*,P_{single}))$, where 
{\small
\begin{gather*}
    MH_{split}(P^*,P_{single}) = \textit{prior ratio} \times \textit{likelihood ratio} \times \textit{proposal ratio}^{-1} \times \lvert \textit{Jacobian} \rvert \nonumber \\
    = \frac{p(L_{\ell_{p},p}^{new}\mid (\sigma_{\ell_p}^2)^{new}, \delta_{\ell_p,p}^{new})p((\sigma_{\ell_p}^2)^{new})p(\delta_{\ell_p,p}^{new})}{p(L_{\ell_{p},p} \mid \sigma_{\ell_p}^2, \delta_{\ell_p,p})p(\sigma_{\ell_p}^2)p(\delta_{\ell_p,p})} \times 
    \frac{p(Y_{\ell_p}\mid \bL_{\ell_p}^{new}, (\sigma_{\ell_p}^2)^{new})}{p(Y_{\ell_p}\mid \bL_{\ell_p}, \sigma_{\ell_p}^2)} \times
    \frac{P-P^*}{P_{single}+1} \times \sqrt{8\sigma_{\ell_p}^2}.
\end{gather*}}
Conversely, when proposing a merge move on one of the $P_{single}$ columns $p$ that contains only one non-zero entry, we first convert column $p$ into a zero column and then apply the reverse transformation of the split move, i.e., $(\sigma_{\ell_p}^2)^{new} = \frac{1}{8}L_{\ell_{p},p}^2 + \sigma_{\ell_p}^2$ and $U = L_{\ell_{p},p}/\sqrt{8L_{\ell_{p},p}^2 + \sigma_{\ell_p}^2}$. The acceptance probability for the merge move is $\min(1,MH_{merge}(P^*,P_{single}))$, which can be computed analogously and satisfies $MH_{merge}(P^*,P_{single})=(MH_{split}(P^*-1,P_{single}-1))^{-1}$.

We carry out posterior inference through the proposed RJMCMC algorithm, with further details provided in the Supplementary Materials Section B.

\section{Simulation Study}
\label{sec:simu}
In this section, we conducted a series of simulation studies to empirically verify our causal identification theory established in Section~\ref{sec:theory}, and evaluate the performance of BayCausal proposed in Section~\ref{sec:mcmc}.

\subsection{Simulation setups} 
We considered two simulation scenarios, each with $n=\{1000,2000,5000\}$ observations. These scenarios were designed to evaluate BayCausal's performance on both directed acyclic graphs (DAGs) and directed cyclic graphs, while incorporating latent confounders. Simulated datasets were generated from the proposed model~\eqref{eqn:full_model}, with 50 replications per scenario and sample size.

\begin{figure}[!htb]
  \centering
    \begin{tabular}{cc}
        \includegraphics[width=0.4\textwidth]{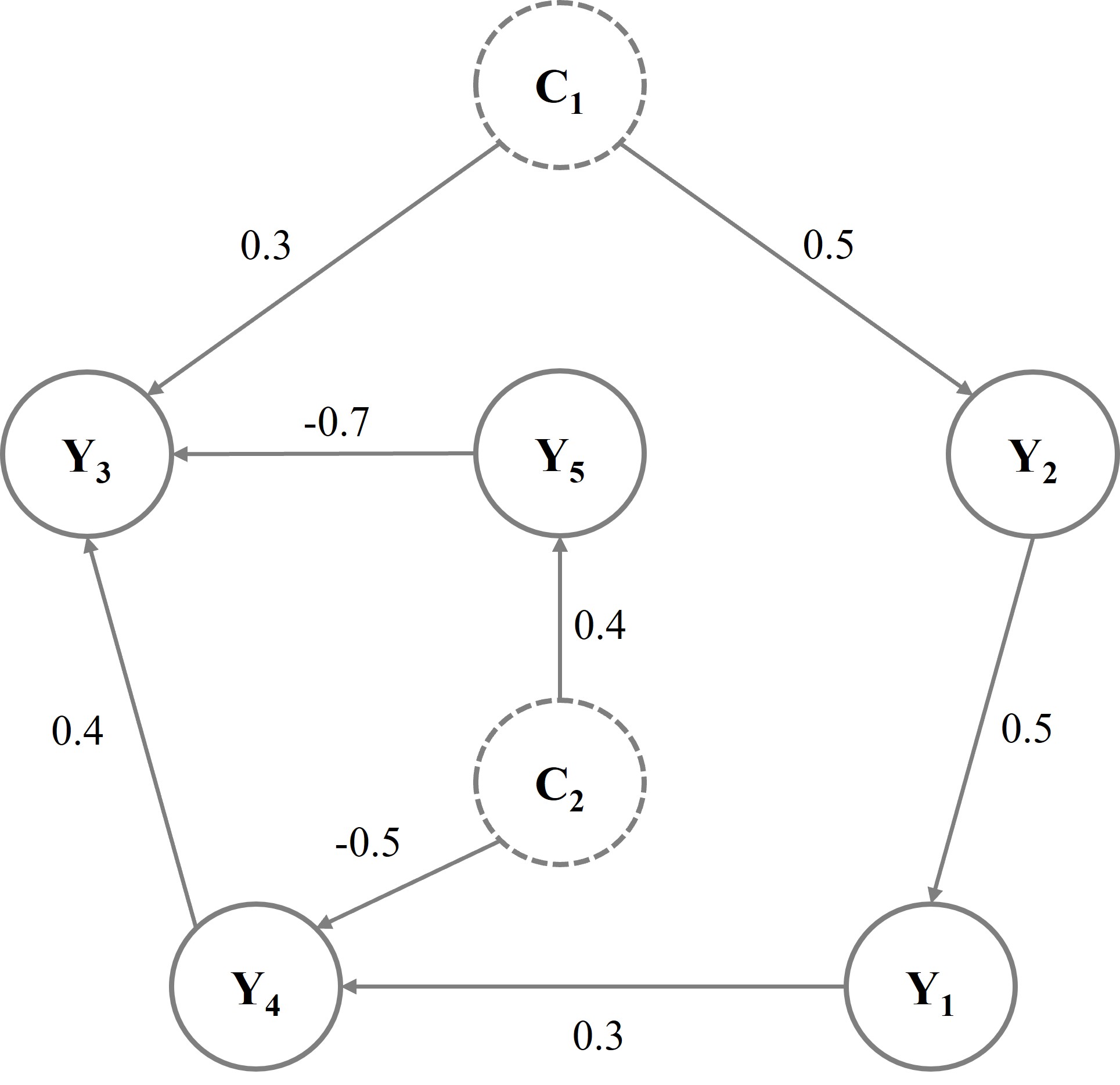} & 
        \includegraphics[width=0.6\textwidth]{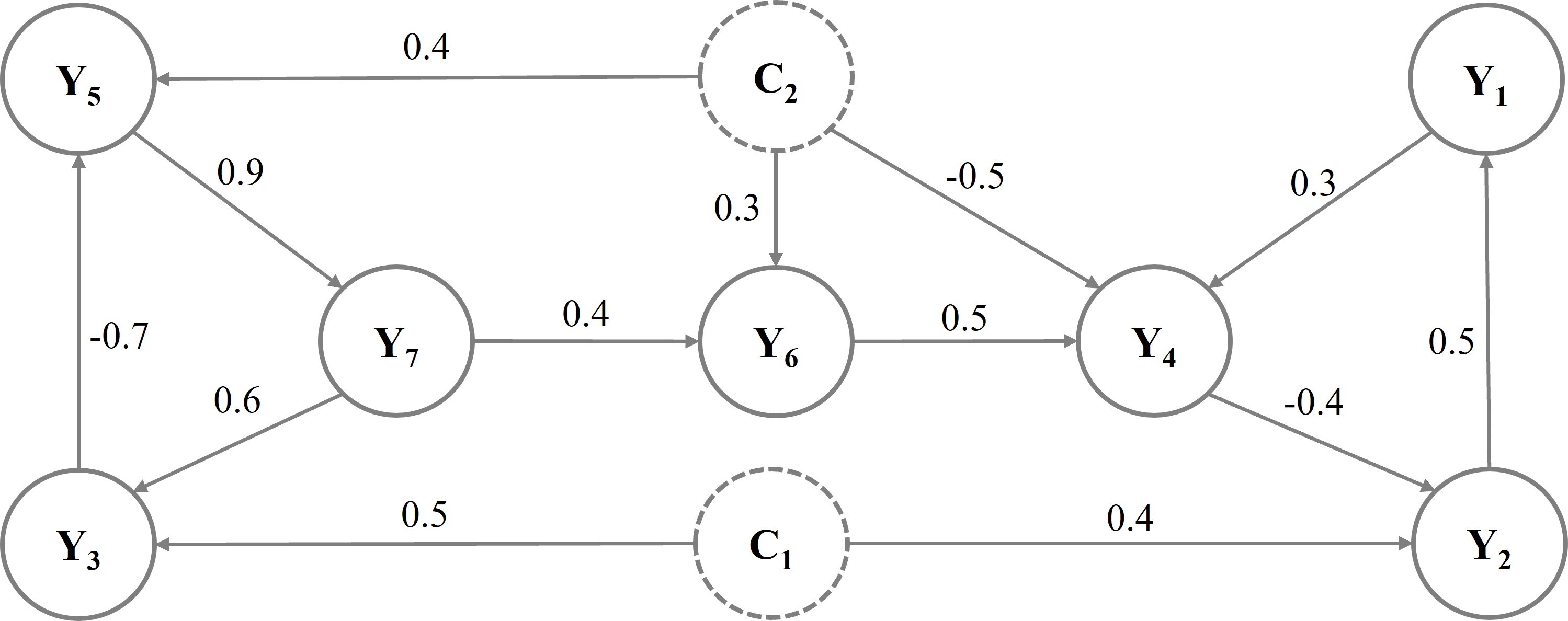} \\
        (a) Scenario I & (b) Scenario II 
    \end{tabular}
   \caption{Simulation truths for Scenarios I and II. Covariates are omitted.} 
   \label{fig:sim}
\end{figure}

In Scenario I, the simulated true causal graph was set to be a DAG consisting of $Q=5$ primary variables and $P^*=2$ latent confounders (Figure~\ref{fig:sim}(a)).   We included $S=2$ covariates generated independently from standard normal distributions, i.e., $\bX=(X_{1},X_2)^{\mT} \sim \mathcal{N}(\bm{0},\bI_2)$. The exogenous errors $\bE$ were independently sampled from Laplace distributions with zero mean and a variance of $0.5$, yielding $\sigma^2_q=1/16$, $q=1,2,\dots,Q$. The non-zero entries in the matrices $\bB$ (i.e., causal relationships among the primary variables) and $\bL$ (i.e., latent confounding effects) were set as follows: $(B_{12},B_{34},B_{35},B_{41})$=$(0.5,0.4,-0.7,0.3)$ and $(L_{21},L_{31},L_{42},L_{52})$=$(0.5,0.3,-0.5,0.4)$. The intercepts were specified as $\bmu$=$(0.79,-0.47$, $-0.26, 0.15, 0.82)^{\mT}$, and the covariate effects were given by $\bA$=$((-0.60,0.80,0.89,0.32,0.26)^{\mT}$, $(-0.88,-0.59,-0.65$, $0.37,-0.23)^{\mT})^{\mT}$. 

In Scenario II, the simulated true causal graph was set to be a directed cyclic graph with $Q=7$ primary variables and $P^*=2$ latent confounders (Figure~\ref{fig:sim}(b)).  It contains two disjoint cycles $Y_1 \to Y_4 \to Y_2 \to Y_1$ and $Y_3 \to Y_5 \to Y_7 \to Y_3$. The covariates $\bX$ and exogenous errors $\bE$ were generated in the same manner as in Scenario I. The non-zero entries in $\bB$ and $\bL$ were set to be $(B_{12},B_{24},B_{53},B_{41},B_{46},B_{75},B_{67},B_{37})$=$(0.5,-0.4,-0.7,0.3,0.5$, $0.9,0.4,0.6)$, and $(L_{21},L_{31},L_{42},L_{52},L_{62})$=$(0.4,0.5,-0.5,0.4,0.3)$. The intercepts were specified as $\bmu=(0.79,-0.47,-0.26,0.15,0.82,-0.60,0.80)$, and the covariate effects were given by $\bA=((0.89,0.32,0.26,-0.88,-0.59,-0.65,0.37)^{\mT},(-0.23,0.54,0,0.44,0.98,-0.24$, $0.55)^{\mT})^{\mT}$. 

We applied BayCausal to the simulated datasets using the following default hyperparameter values: $a_{\nu}=b_{\nu}=1$, $a_{\rho}=b_{\rho}=1$, $a_{\sigma}=b_{\sigma}=1$, $a_{\kappa}=b_{\kappa}=1$, and $\nu_{0}=2.5\times10^{-4}$. We set the maximum number of latent confounders $P$ to be $Q-1$. For the hyperparameters associated with latent confounders, we chose $b_1=b_2=6$ and $c_1=c_2=6(P-1)/P$, following  \cite{fruhwirth2024sparse}. We ran 50,000 MCMC iterations with a burn-in of 30,000 iterations and a thinning factor of 10 for each analysis. To determine whether the estimated causal effects are zeros or non-zeros, we employed the median probability model criteria \citep{barbieri2004optimal}. Specifically, an edge $Y_{q'} \to Y_q$ was included in the estimated graph if  $P(\gamma_{qq'}=1) > 0.5$ calculated from the post-burn-in MCMC samples. The same rule was used to determine the existence of $L_p \to Y_q$ and $X_s \to Y_{q}$.

For comparison, we considered three state-of-the-art causal discovery methods as alternatives. The first method is BFCI \citep{andrews2023fast}, an improved version of the Greedy Fast Causal Inference (GFCI; \citealt{ogarrio2016hybrid}) algorithm. GFCI combines the strengths of both constraint-based and score-based approaches and has been shown to outperform standard constraint-based algorithms such as FCI \citep{spirtes1995causal} and RFCI \citep{colombo2012learning}. BFCI further improves upon the GFCI algorithm by incorporating the best order score search algorithm in its initial step, leading to more accurate causal graph estimations. Although BFCI accounts for latent confounders, it is restricted to DAGs within the Markov equivalence class and cannot handle cycles. Furthermore, BFCI outputs only the causal structure without providing estimates of causal effects. The second method is Repetitive Causal Discovery (RCD; \citealt{maeda2020rcd}), a model-based algorithm that repeatedly infers causal directions between small subsets of observed variables and determines whether their relationships are influenced by latent confounders. RCD is limited to identifying causal effects among observed variables in DAGs and does not estimate the latent confounding effects. The third method is the Linear Non-Gaussian Discovery (LiNG-D; \citealt{lacerda2008discovering}) algorithm, a model-based approach that accommodates cyclic causal structures but does not allow for latent confounders. As none of the three alternative methods account for covariates, we generated a separate corresponding dataset for each repeated experiment by removing the covariates from the data-generating process of the proposed model~\eqref{eqn:full_model} to ensure a fair comparison. We then implemented the three alternative methods using the Python packages \pkg{py-tetrad} \citep{ramsey2023py} and \pkg{causal-learn} \citep{zheng2024causal}, with their default settings. 

Lastly, we conducted two additional simulation studies to evaluate the robustness of BayCausal under model misspecification.  First, we replaced the Laplace-distributed error terms $E_q$ with a Student's $t$-distribution to test sensitivity to the form of non-Gaussianity. Second, we substituted the Gaussian latent confounders $C_p$ with Laplace-distributed variables to assess the impact of violating the Gaussian confounder assumption. All other settings remained identical to Scenarios I and II, using a sample size of $n=5000$. 

\subsection{Simulation results} 
We first evaluated the performance of BayCausal in recovering the causal structure among the primary variables using the ground truth matrix $\bB$ as a reference. The following two evaluation metrics were computed. The first was an exact-match criterion: the count of successful recoveries (CSR), defined as the number of repetitions in which the estimated directed graph over the primary variables matched the true graph implied by $\bB$ exactly. The second comprised edge-wise accuracy metrics computed over all ordered pairs of distinct primary variables: the true positive rate TPR = TP/(TP + FN), the false discovery rate FDR = FP/(TP + FP), and the Matthews correlation coefficient $\text{MCC}=(\text{TP} \times \text{TN} - \text{FP} \times \text{FN})/\sqrt{(\text{TP}+\text{FP})(\text{TP}+\text{FN})(\text{TN}+\text{FP})(\text{TN}+\text{FN})}$, where TP, TN, FP, and FN denote true positives, true negatives, false positives, and false negatives, respectively.   MCC ranges from -1 (perfect disagreement) through 0 (no better than random guessing) to 1 (perfect agreement). All metrics were computed for the graph estimates produced by BayCausal and all alternative methods.

Table~\ref{tab:simu} summarizes results for Scenarios I and II across varying sample sizes. In Scenario I, performance improved with sample size for all methods, but BayCausal consistently outperformed the alternatives at every $n$.  Specifically, at $n=5000$, BayCausal successfully recovered the true underlying causal graph in all 50 repeated experiments, achieving perfect performance with a TPR of 1.00, FDR of 0.00, and MCC of 1.00. In contrast, BFCI and RCD achieved exact recovery in 7 and 6 out of 50 repetitions, respectively, while LiNG-D did not recover the correct graph in any repetition, consistent with its inability to handle latent confounding. The MCC values for BFCI, RCD, and LiNG-D were 0.71, 0.85, and 0.69, respectively, which were substantially lower than the perfect MCC of 1.00 achieved by BayCausal. In Scenario II, none of the alternative methods achieved exact recovery in any of the 50 repetitions, whereas BayCausal maintained a high CSR and reached perfect reconstruction at $n=5000$. These results highlight BayCausal's advantage in jointly accommodating cyclic causal structures and latent confounding.

\begin{table}[!htb]
  \centering
  \small
  \begin{tabular}{llrrrrrrrrr}
    \toprule
    & & \multicolumn{4}{c}{Scenario I} & \multicolumn{4}{c}{Scenario II} \\
    \cmidrule(lr){3-6} \cmidrule(lr){7-10}
    Sample Size & Method & CSR & TPR & FDR & MCC & CSR & TPR & FDR & MCC \\
    \midrule
    \multirow{4}{*}{$n = 1000$}
      & BayCausal   & 27 & 1.00 & 0.13 & 0.92 & 43 & 0.99 & 0.04 & 0.97 \\
      & BFCI       & 1 & 0.61 & 0.41 & 0.52 & 0 & 0.40 & 0.51 & 0.35 \\
      & RCD        & 3 & 0.66 & 0.20 & 0.68 & 0 & 0.41 & 0.45 & 0.39 \\
      & \text{LiNG-D} & 0 & 1.00 & 0.49 & 0.65 & 0 & 1.00 & 0.50 & 0.63 \\
      \hline
    \addlinespace[2pt]
    \multirow{4}{*}{$n = 2000$}
      & BayCausal   & 47 & 1.00 & 0.01 & 0.99 & 49 & 1.00 & 0.00 & 1.00 \\
      & BFCI       & 3 & 0.64 & 0.39 & 0.55 & 0 & 0.43 & 0.54 & 0.34 \\
      & RCD        & 6 & 0.92 & 0.27 & 0.78 & 0 & 0.51 & 0.56 & 0.37 \\
      & \text{LiNG-D} & 0 & 1.00 & 0.44 & 0.69 & 0 & 1.00 & 0.44 & 0.69 \\
      \hline
    \addlinespace[2pt]
    \multirow{4}{*}{$n = 5000$}
      & BayCausal & 50 & 1.00 & 0.00 & 1.00 & 50 & 1.00 & 0.00 & 1.00 \\
      & BFCI     & 7 & 0.79 & 0.27 & 0.71 & 0 & 0.71 & 0.26 & 0.67 \\
      & RCD      & 6 & 0.98 & 0.22 & 0.85 & 0 & 0.34 & 0.70 & 0.18 \\
      & \text{LiNG-D} & 0 & 1.00 & 0.43 & 0.69 & 0 & 1.00 & 0.45 & 0.68 \\
    \addlinespace[2pt]
    \bottomrule
  \end{tabular}
  \caption{Performance comparisons of the proposed method (BayCausal) and alternative methods (BFCI, RCD, and LiNG-D) in Scenarios I and II across varying sample sizes. Metrics summarized over 50 runs: CSR (count of successful recoveries), TPR (true positive rate), FDR (false discovery rate), and MCC (Matthews correlation coefficient).}
  \label{tab:simu}
\end{table}

\begin{figure}[!htb]
  \centering
    \begin{tabular}{cc}
       \includegraphics[width=0.4\textwidth]{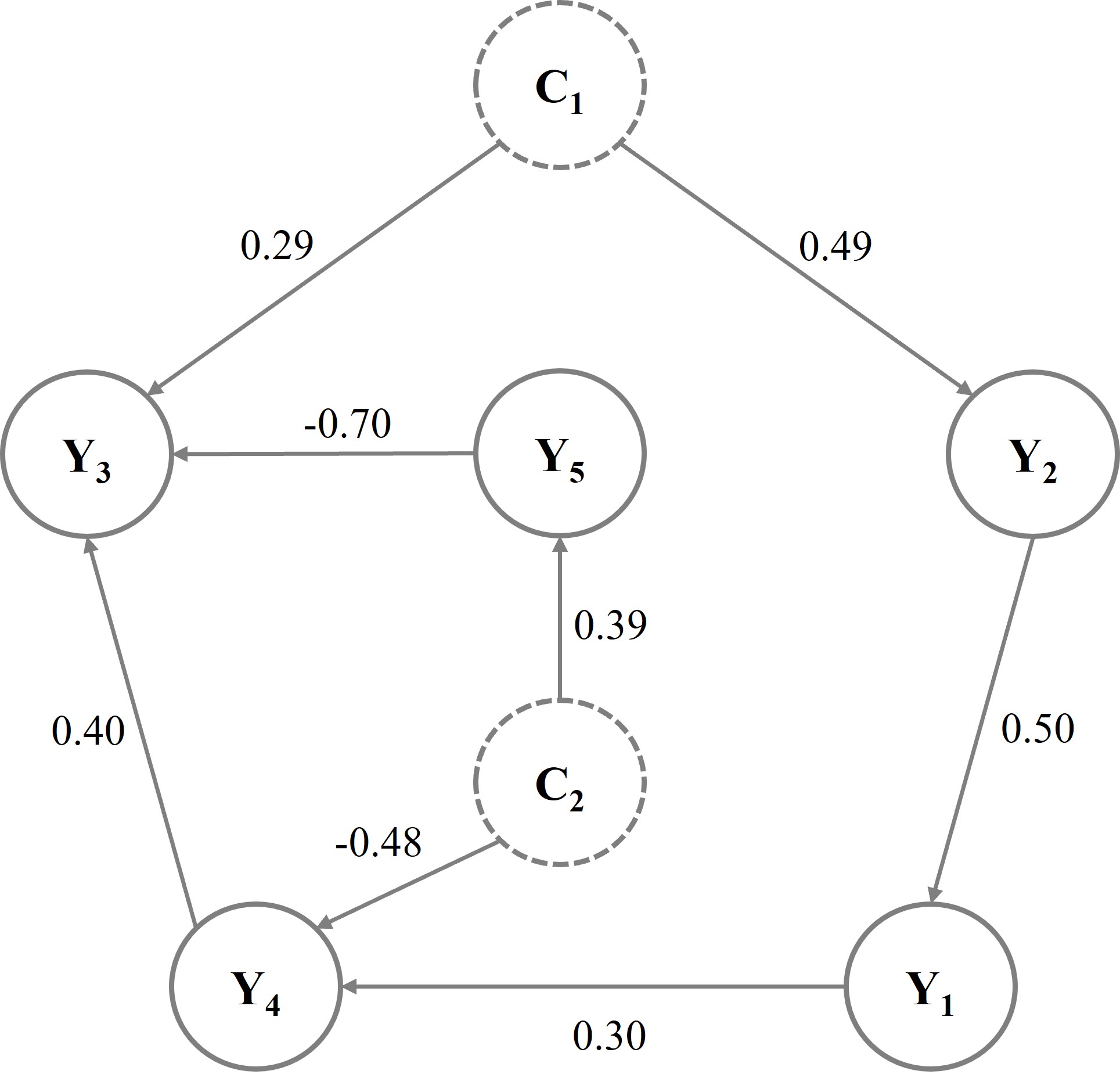} &
       \includegraphics[width=0.4\textwidth]{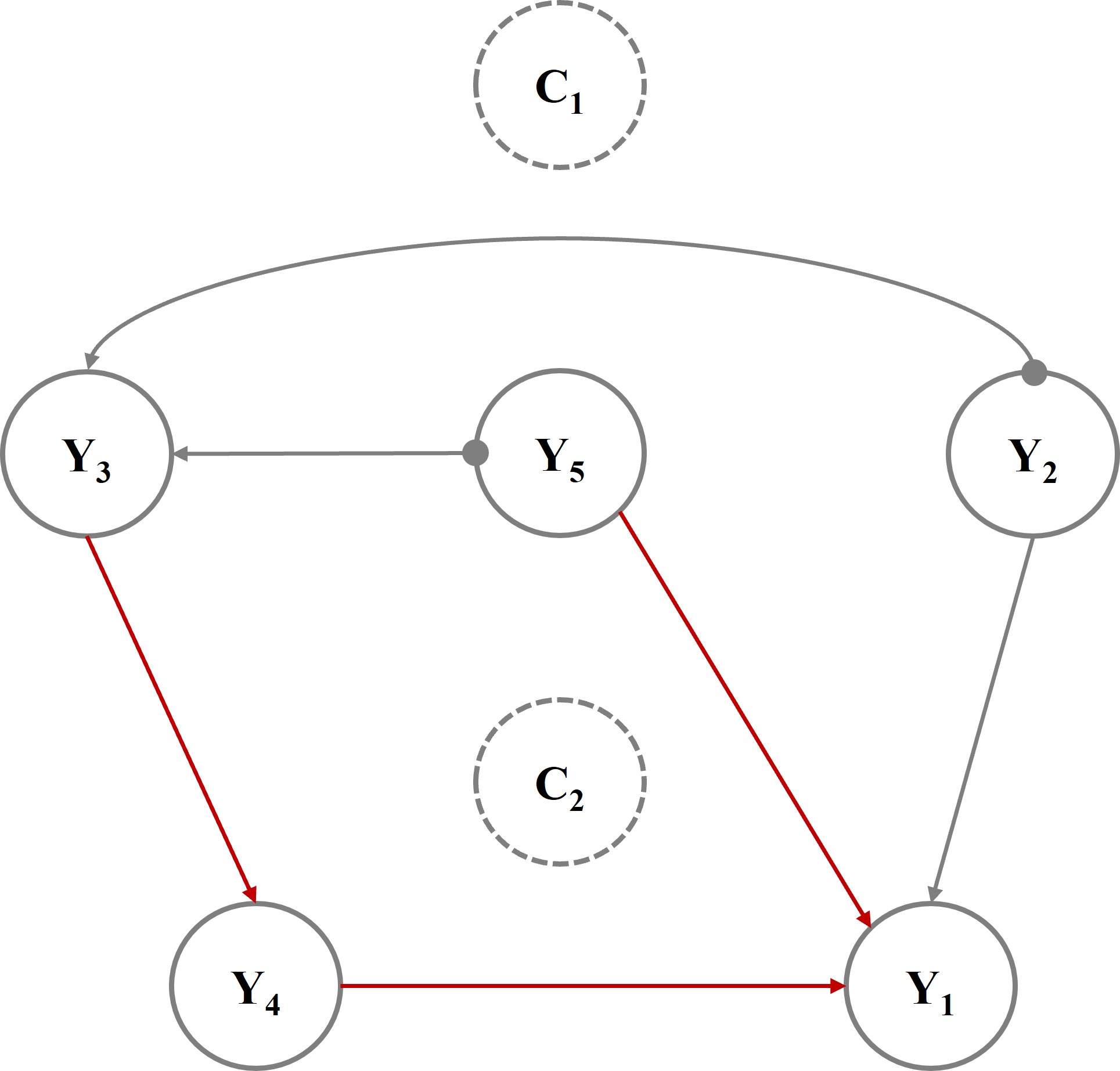} \\
       (a) $\widehat{\mG}_{\text{BayCausal}}$ & (b) $\widehat{\mG}_{\text{BFCI}}$ 
    \end{tabular}
    \begin{tabular}{cc}
       \includegraphics[width=0.4\textwidth]{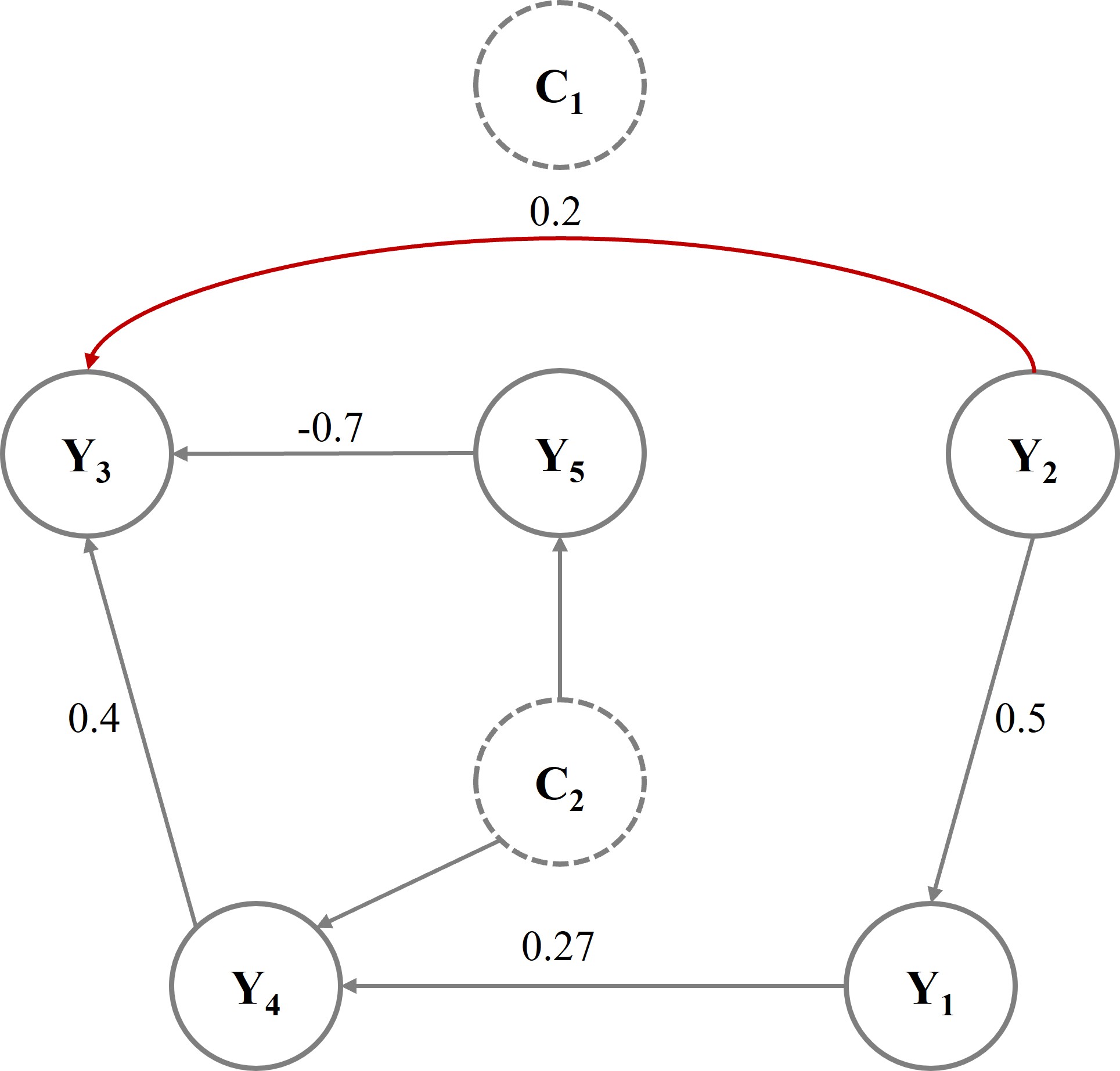} &
       \includegraphics[width=0.4\textwidth]{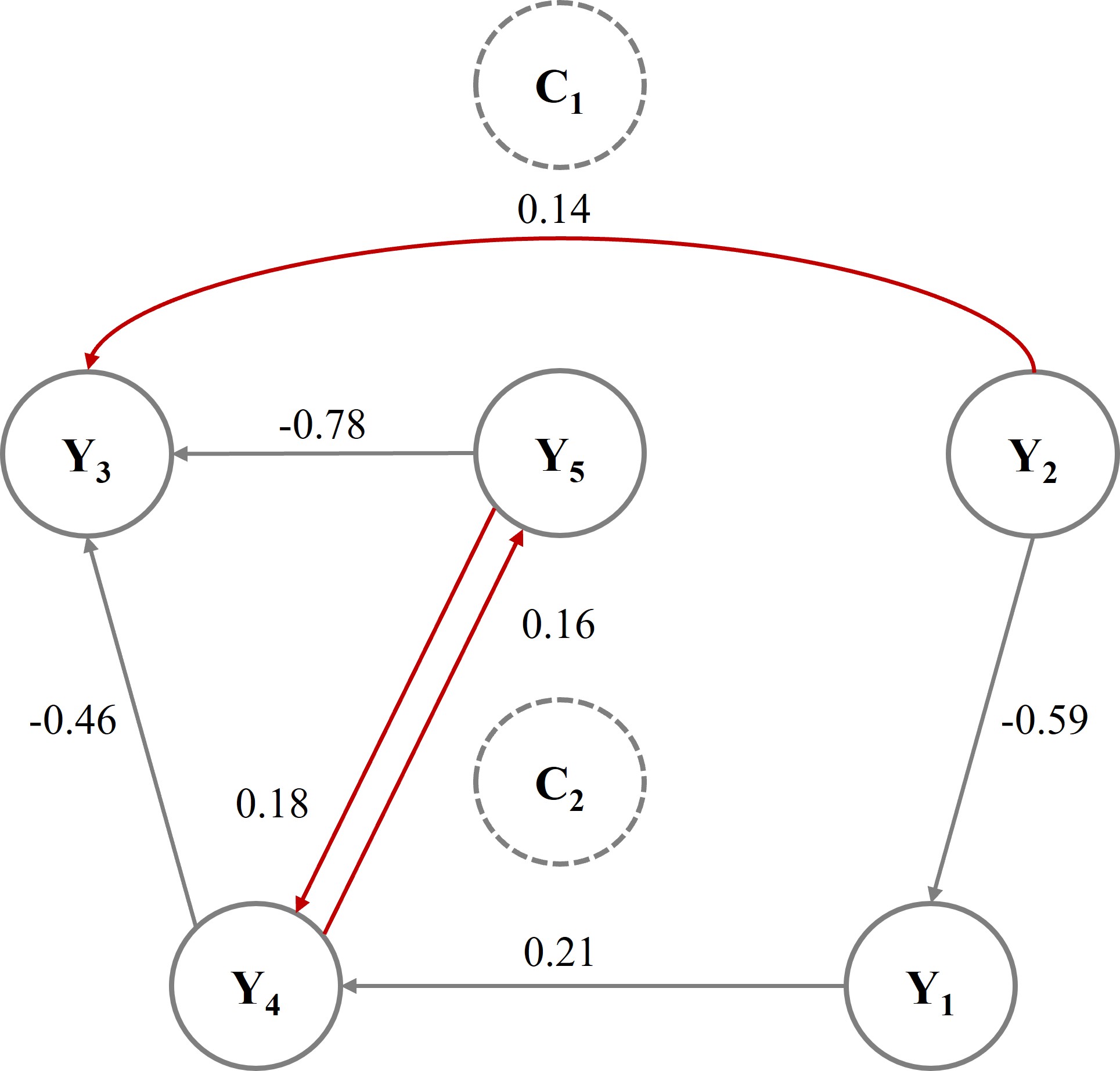} \\
       (c) $\widehat{\mG}_{\text{RCD}}$ & (d) $\widehat{\mG}_{\text{LiNG-D}}$ 
    \end{tabular}
   \caption{The estimated graph modes under BayCausal and alternative methods ($\text{BFCI}$, $\text{RCD}$, and \text{LiNG-D}) for Scenario I. The red lines indicate inconsistent edges between the estimated and true causal graphs. Covariates are omitted from all graphs.} 
   \label{fig:sim1}
\end{figure}

\begin{figure}[!htb]
  \centering
    \begin{tabular}{cc}
       \includegraphics[width=0.5\textwidth]{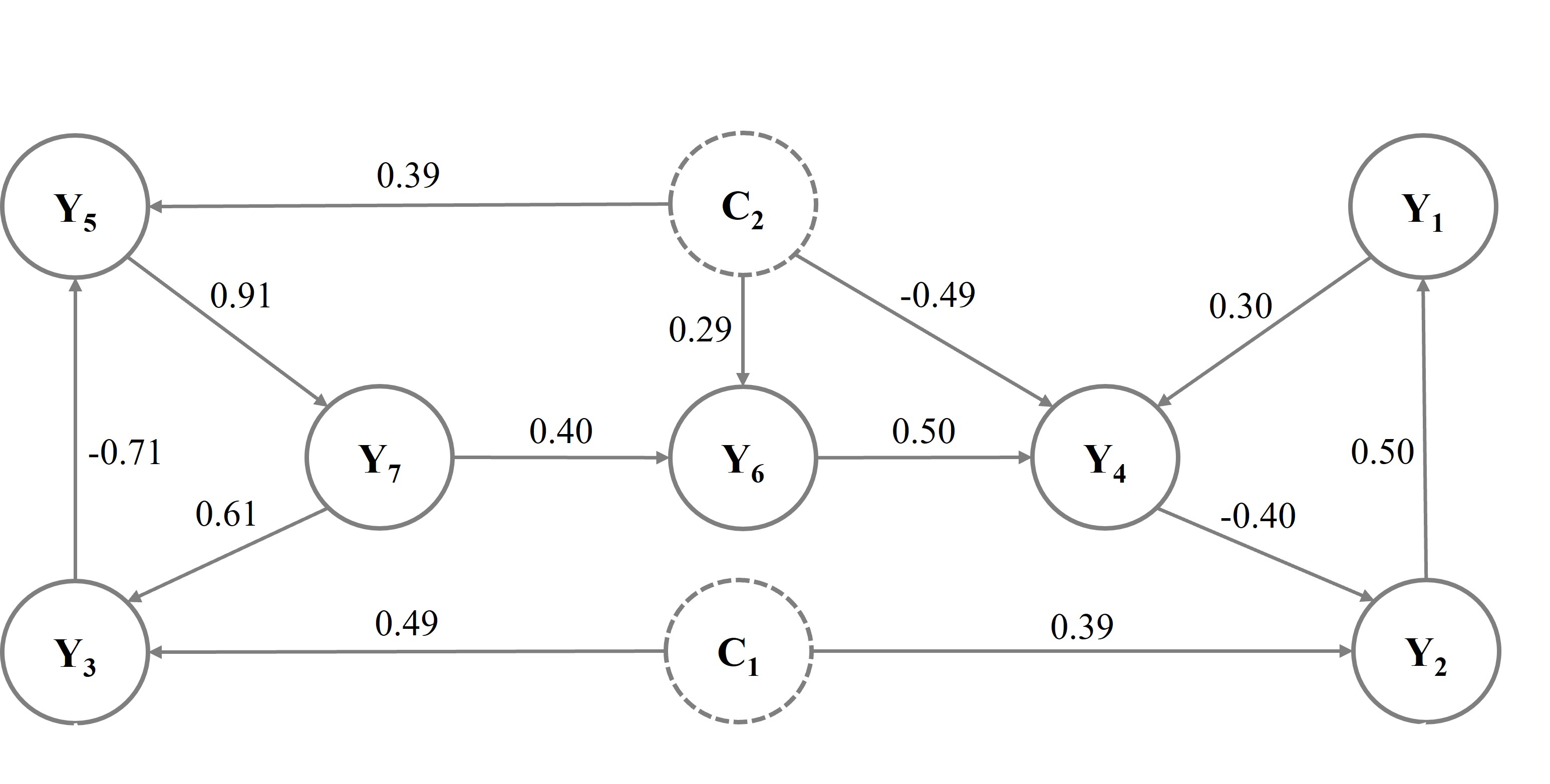} &
       \includegraphics[width=0.5\textwidth]{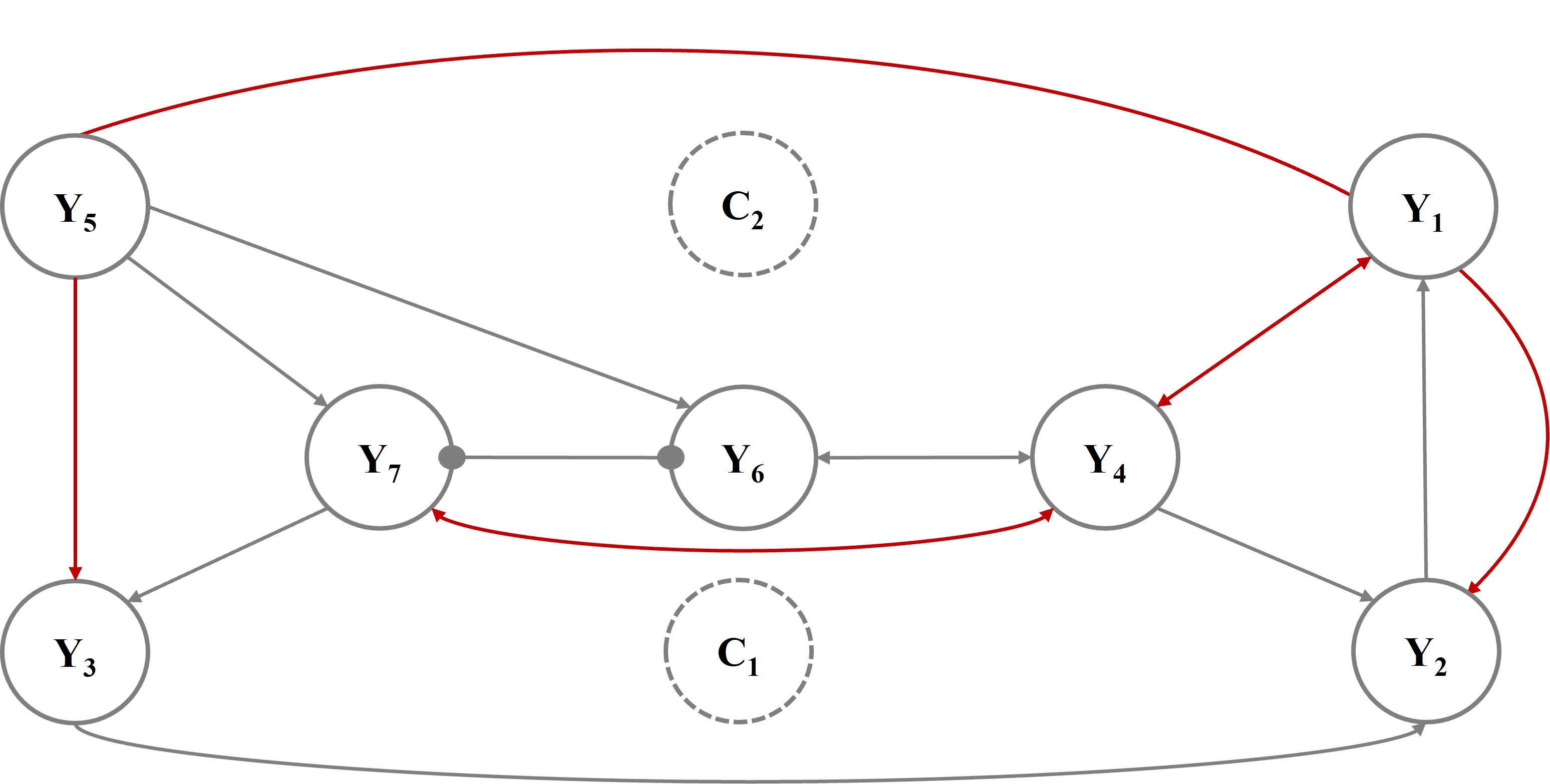} \\
       (a) $\widehat{\mG}_{\text{BayCausal}}$ & (b) $\widehat{\mG}_{\text{BFCI}}$ 
    \end{tabular}
    \begin{tabular}{cc}
       \includegraphics[width=0.5\textwidth]{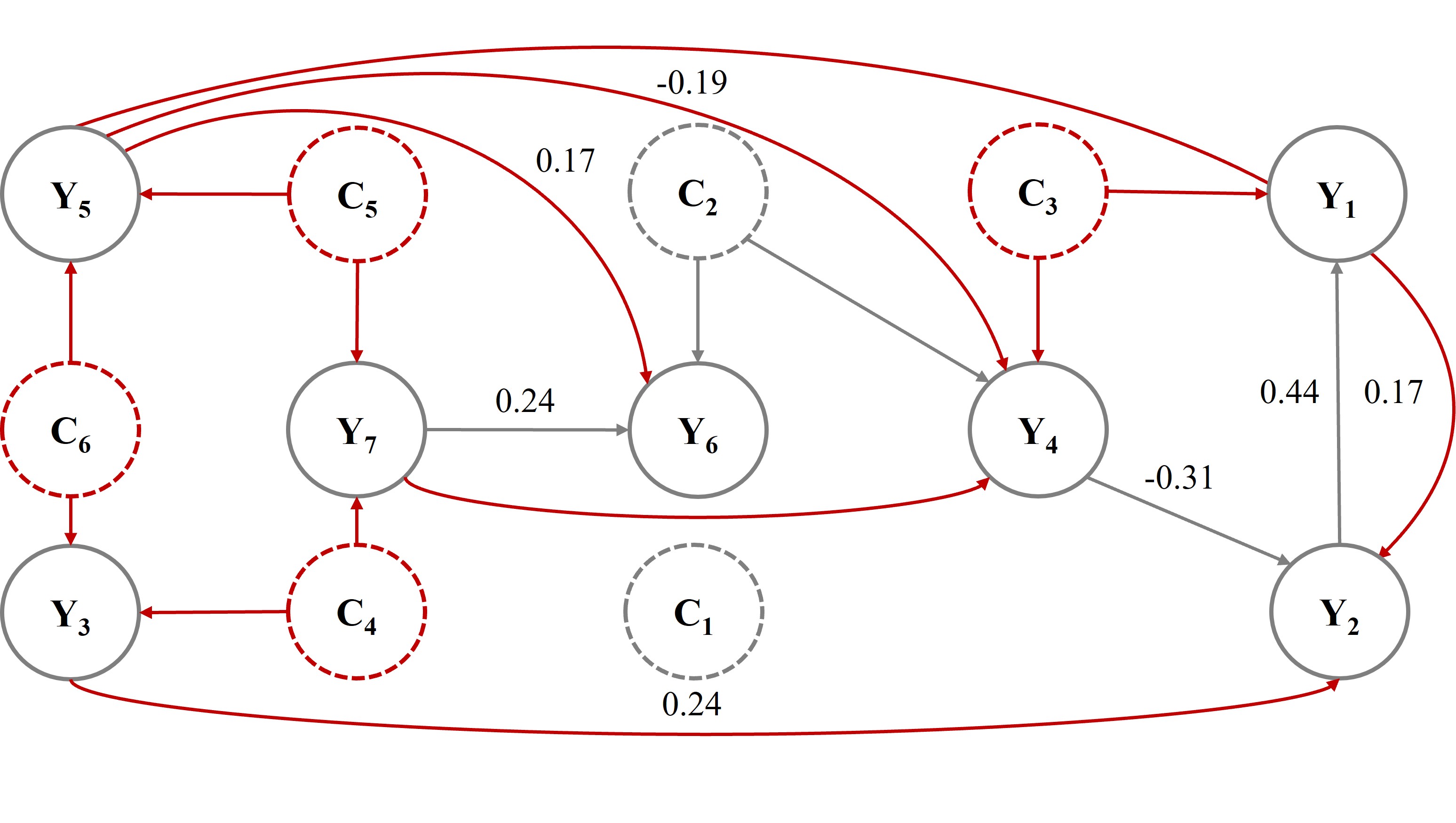} &
       \includegraphics[width=0.5\textwidth]{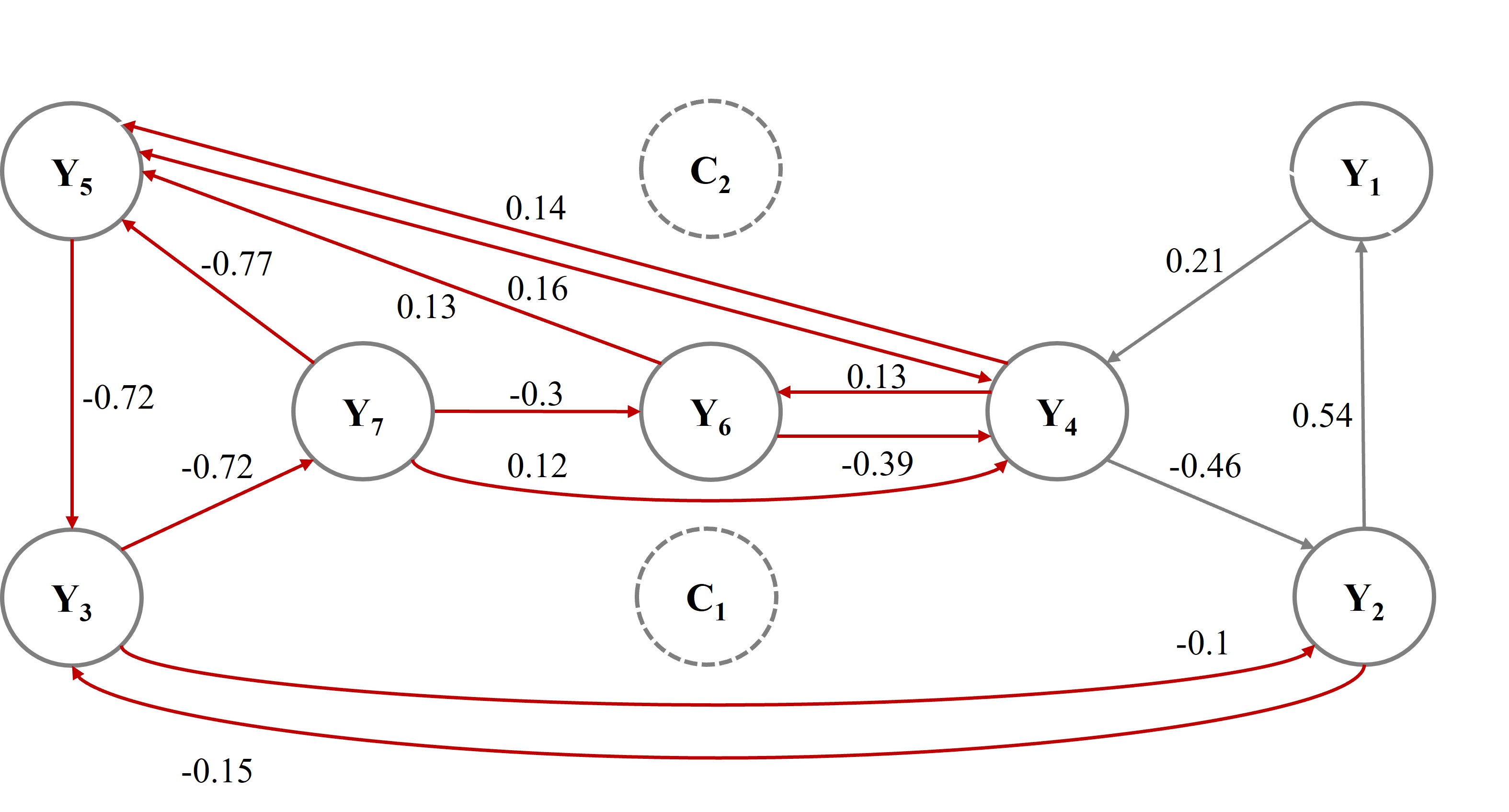} \\
       (c) $\widehat{\mG}_{\text{RCD}}$ & (d) $\widehat{\mG}_{\text{LiNG-D}}$ 
    \end{tabular}
   \caption{The estimated graph modes under BayCausal and alternative methods ($\text{BFCI}$, $\text{RCD}$, and \text{LiNG-D}) for Scenario II. The red lines indicate inconsistent edges between the estimated and true causal graphs. Covariates are omitted from all graphs.} 
   \label{fig:sim2}
\end{figure}

Next, we evaluated the effectiveness of all methods in recovering causal structures by examining their estimated causal graphs, including both the primary variables and the latent confounders. For each method and scenario, we identified the graph mode, defined as the most frequently observed graph across 50 replicates. Figures~\ref{fig:sim1}(a)–(d) and \ref{fig:sim2}(a)–(d) display the mode graphs for $n = 5000$ under Scenario I and Scenario II, respectively. Causal-effect estimates were summarized by averaging the estimates associated with the mode graph for each method.

As shown in Figures~\ref{fig:sim1}(a) and \ref{fig:sim2}(a), BayCausal not only recovered the underlying causal structures but also achieved high accuracy in estimating both causal effects and latent confounding effects.  For example, in Scenario I, the latent confounder effects from $C_1$ to $Y_2$ and $Y_3$, with true values of 0.5 and 0.3, were accurately estimated by BayCausal as 0.49 and 0.29, respectively. In Scenario II, BayCausal successfully identified the cyclic relationships $Y_1 \to Y_4 \to Y_2 \to Y_1$ and accurately recovered the corresponding effects (0.3, -0.4, 0.5). 

In Scenario I, the graph returned by BFCI is a partial ancestral graph (PAG), which compactly encodes the Markov equivalence class of causal models compatible with the data when latent confounding may be present. In a PAG, edge endpoints convey what is guaranteed versus what remains possible: a tail ``--'' marks ancestral (causal) direction at that end, an arrowhead ``$>$'' rules out ancestry at that end, and a circle ``$\bullet/\circ$'' denotes that the orientation at that end is not identified. Supplementary Table~S1 summarizes the semantics used here. For instance, BFCI produced a $Y_2~\bullet\!\rightarrow~Y_3$ edge, which in a PAG rules out $Y_3$ causing $Y_2$ and leaves open that $Y_2$ causes $Y_3$, or that $Y_2$ and $Y_3$ share an unmeasured confounder, or both. This is consistent with the true latent confounder $C_1$, but not conclusive on its own.  By contrast, BFCI did not indicate latent confounding between $Y_4$ and $Y_5$ (e.g., via a bidirected $Y_4 \leftrightarrow Y_5$ edge), and thus missed $C_2$. RCD successfully identified the latent confounder $C_2$ between $Y_4$ and $Y_5$, and correctly inferred nearly all of the causal effects. However, it mistakenly interpreted the latent confounding between $Y_2$ and $Y_3$ as a direct causal relationship. Since LiNG-D does not account for latent confounders, it incorrectly represented relationships between primary variables as either direct causal links (e.g., $Y_2\to Y_3$) or cyclic dependencies (e.g., $Y_4 \rightleftarrows Y_5$) whenever latent confounders were actually present. 

In Scenario II, where both cycles and latent confounders are present, all alternative methods exhibited substantial discrepancies from the true graph: none was able to simultaneously recover cyclic causal relationships and latent confounding. These results further highlight the advantages of BayCausal, which is uniquely designed to address both challenges with high accuracy. 
 
Furthermore, we evaluated the performance of BayCausal in estimating the covariate effects and the number of latent confounders, both of which are its unique strengths compared to all alternative methods. Supplemental Tables S2 and S3 report summary statistics for Scenario I and Scenario II at $n=5000$, including the mean and standard deviation of the posterior means across 50 replicates, the coverage of the 95\% credible intervals, and the bias and mean squared error (MSE). These results show that BayCausal accurately estimates the covariate effects. Supplementary Figure S2 presents histograms of the estimated number of latent confounders in Scenarios I and II; in both cases, the distributions are centered at the simulated true value $P^{*}=2$ with high probability. 

Lastly, Supplementary Table S4 reports the sensitivity-analysis results under model misspecification. Across two experimental settings, BayCausal consistently outperformed all alternatives. In the first setting, which varied the non-Gaussian error distributions, it achieved perfect graph recovery, indicating strong robustness to the choice of error distribution. In the second setting, even when the identifiability assumption that latent confounders are Gaussian was violated, the method remained robust, yielding near-perfect graph recovery (e.g., MCC 0.92 in Scenario I and 0.99 in Scenario II).

\section{Application: WIHS Data Analysis}
\label{sec:wihs}
The Women's Interagency HIV Study (WIHS; \citealt{adimora2018cohort}) is a large, prospective, observational cohort study designed to investigate the long-term health outcomes including comorbidities among women with and at risk for HIV. Participants' sociodemographic, clinical, and behavioral data were collected through clinical interviews, self-reported surveys, and laboratory assessments. For the current analysis, we focused on women living with HIV from the Washington, D.C. site and aimed to examine causal relationships among $Q = 8$ health outcomes that are routinely assessed during treatment decision-making. 
These outcomes included depression measured by the Center for Epidemiological Studies Depression Scale (CES-D; \citealt{radloff1977ces}), spanning somatic symptoms (Somatic) such as poor appetite and restless sleep, negative affect (Negative) such as sadness, lack of positive affect (Positive) such as hopelessness, and interpersonal symptoms (Interpersonal) such as feeling people dislike me; HIV RNA viral load (Vload); CD4 count (CD4N); estimated glomerular filtration rate (eGFR), an indicator for kidney function; and body mass index (BMI). To empirically validate the non-Gaussianity assumption of our causal identification theory (Assumption~\ref{asm:causal}(b)), we performed Shapiro-Wilk tests \citep{shaphiro1965analysis} on all primary variables. The null hypothesis of normality was rejected in all cases, thereby satisfying the theoretical requirement.  We also considered the following covariates: age, race, education, and diabetes. Our analysis included all available observations from all participants, resulting in a total sample size of 6,777 for this study. The extension of the BayCausal method to model longitudinal data is an important direction for future work and will be discussed in Section~\ref{sec:con}.

\begin{figure}[!htb]
  \centering
  \includegraphics[width=\textwidth]{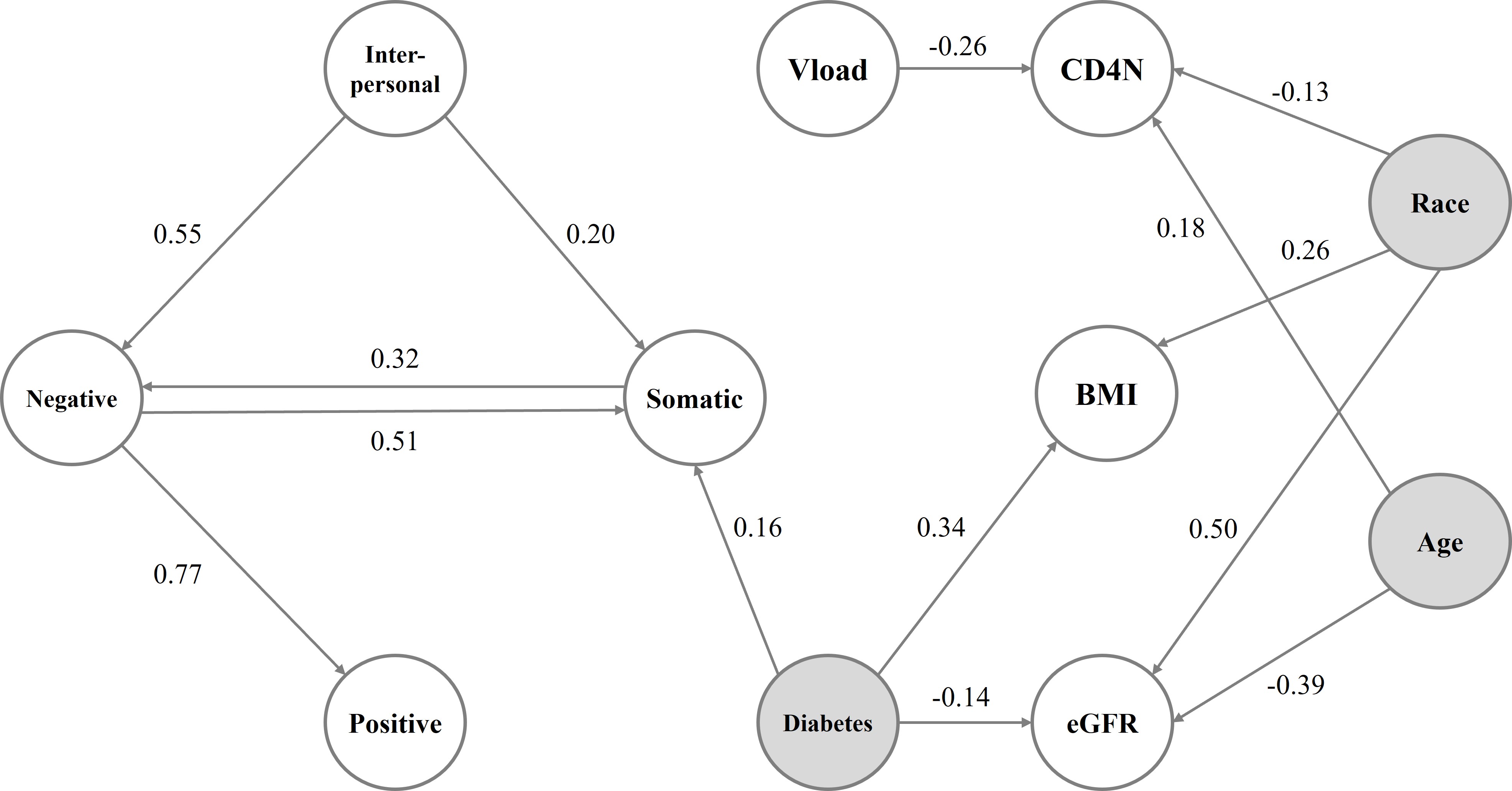}
  \caption{Estimated causal graph under BayCausal for the WIHS data analysis. The white and gray circles indicate the primary variables and covariates, respectively.}
  \label{fig:real_data}
\end{figure}

We applied BayCausal to the WIHS data using the same set of hyperparameters as in the simulation studies. Five independent MCMC chains with different initial values were run, each with 100,000 iterations, a burn-in of 80,000 iterations, and a thinning factor of 10, for posterior inference. Convergence diagnostics confirmed that all MCMC chains converged to the same stationary distribution. Figure~\ref{fig:real_data} presents the estimated causal graph under BayCausal, summarized using the median probability model criteria as in the simulation study.

As shown in Figure~\ref{fig:real_data}, BayCausal identified a direct negative causal effect of viral load on CD4 count. This aligns with established clinical knowledge, as HIV primarily targets CD4 cells, and viral replication directly leads to their depletion \citep{vidya2017pathophysiology}. Furthermore, the model identified a direct negative effect of age on eGFR. This is consistent with the known physiology of aging, which involves a gradual decline in kidney function \citep{weinstein2010aging}.

BayCausal also revealed a network of causal relationships among the four depression-related variables. Notably, a cyclical relationship was identified between somatic symptoms and negative affect. This suggests a reinforcing feedback loop where an increase in somatic symptoms elevates negative affect, which in turn exacerbates the physical symptoms. This finding is supported by existing literature; for instance, \cite{leventhal1997does} linked stress and illness to increased negative affect, while \cite{charles2006daily} demonstrated that negative affect is a risk factor for adverse health outcomes. Additionally, negative affect was found to directly cause a lack of positive affect, a finding consistent with prior research on the interdependence of affective states \citep{diener1984independence}. These relationships highlight how mental and physical well-being are dynamically interconnected in people with HIV.

\section{Conclusion}
\label{sec:con}
We proposed a novel framework for causal discovery in the presence of both cycles and latent confounders. By leveraging identifiability results from noisy ICA and FA, we demonstrated that both the causal relationships among the primary variables and the causal structures of the latent confounders can be uniquely identifiable from observational data. Building on this theoretical foundation, we developed an RJMCMC algorithm, BayCausal, for Bayesian causal structure learning and showed its superior performance compared with state-of-the-art alternatives through extensive simulation studies. An application to a real-world HIV database illustrated the practical utility of BayCausal in uncovering clinically meaningful relationships among comorbidities in people living with HIV. Currently, no existing publicly available software for causal discovery achieves unique identifiability while accounting for both cycles and latent confounders. We hope that our R package, \pkg{BayCausal}, will become a standard computational tool to fill this important gap. 

Several promising directions remain for future work. First, the unique causal identifiability of the proposed framework relies on the condition that the directed graphs contain only disjoint cycles. The theoretical framework can potentially be extended to accommodate longitudinal data and more general cyclic structures 
when additional instrumental variables are available \citep{jin2025directed}. We can extend BayCausal to capture time-lagged causality among primary variables using a vector autoregression model \citep{swanson1997impulse}. Imposing structural constraints such as low-rank representations on the covariance matrix of time-varying latent confounders may help ensure identifiability in the longitudinal setting by simplifying their cross-time correlations. Second, the proposed framework is based on a linear additive data-generating model. Extending it to more flexible frameworks, such as nonlinear additive models or functional causal models, would be an interesting future direction. Lastly, BayCausal may be computationally intensive in high-dimensional settings. Developing more computationally efficient algorithms for causal structure learning could enhance the scalability and practical applicability of the proposed framework.

\section*{Acknowledgments}
This work is supported in part by National Institute of Health (NIH) grants R01MH128085 and R01AI197147. The contents of this publication are solely the responsibility of the authors and do not represent the official views of the NIH. MWCCS (Principal Investigators): Data Analysis and Coordination Center (Gypsyamber D'Souza, Stephen Gange, and Elizabeth Topper); Metropolitan Washington CRS (Seble Kassaye and Daniel Merenstein).

\clearpage
\begin{center}
\textbf{\Large Supplemental Materials for ``Bayesian Causal Discovery with Cycles and Latent Confounders''}
\end{center}
\beginsupplement

\section*{A. Technical Proofs}

\subsection*{A1. Proof of Lemma~1}

\begin{proof}
We begin the proof by first showing that $\bI - \bB$ is invertible given that $\bB$ is stable. Specifically, the stability condition indicates that the maximum modulus of $\bB$'s eigenvalues is strictly less than 1. By Theorem 5.6.12 in \cite{horn2012matrix}, there exists a matrix norm $\|\cdot\|$ such that $\|\bB\| < 1$. Then the series $\sum_{n=0}^{\infty}\bB^n$ converges. By Corollary 5.6.16 in \cite{horn2012matrix}, since $(\bI - \bB)\sum_{n=0}^{N}\bB^n=\bI-\bB^{N+1}\to \bI$ as $N\to \infty$, we conclude that $\bI - \bB$ is invertible and $(\bI - \bB)^{-1}=\sum_{n=0}^{\infty}\bB^n$ is invertible. We define $\widetilde{\bB} = (\bI-\bB)^{-1}$. 

Next, we show that the matrix $\widetilde{\bB}$ is identifiable up to permutation and rescaling. Rewrite model
\begin{equation*}
    \bY = \bB\bY + \bL\bC + \bE,
\end{equation*}
in the manuscript as:
\begin{equation*}
\bY = (\bI-\bB)^{-1}(\bL\bC + \bE) = \widetilde{\bB}(\bL\bC + \bE) = \begin{bmatrix} \widetilde{\bB},\,\widetilde{\bB}\bL \end{bmatrix}
\begin{bmatrix} \bE \\ \bC \end{bmatrix} = \begin{bmatrix} \widetilde{\bB},\, \widetilde{\bL} \end{bmatrix} \begin{bmatrix} \bE \\ \bC \end{bmatrix},
\end{equation*}
where $\widetilde{\bL} = \widetilde{\bB}\bL$. Suppose that we have an equivalent representation
\begin{equation*}
\bY = \begin{bmatrix} \widetilde{\bB}',\, \widetilde{\bL}' \end{bmatrix} \begin{bmatrix} \bE' \\ \bC' \end{bmatrix}
\end{equation*}
where $\bE'$ and $\bC'$ denote the independent non-Gaussian and Gaussian sources, respectively. Then by Theorem 10.3.1 in \cite{kagan1973characterization} and Theorem A 6.1 in \cite{eriksson2003identifiability}, every column of $\widetilde{\bB}$ is a multiple of a column of $\widetilde{\bB}'$ and vice versa. Therefore, $\widetilde{\bB}$ is identifiable up to column permutation and rescaling. 

Lastly, we show that there exists a unique permutation and scaling corresponding to the graphical representation $\mG$. We first consider the case where $\mG$ is a directed acyclic graph (DAG), and then extend the argument to directed graphs with disjoint cycles. 

Let $\bW=\widetilde{\bB}^{-1}=\bI-\bB$. When the underlying true graph $\mG$ is a DAG, Appendix A of \cite{shimizu2006linear_supp} demonstrates that there exists a unique column permutation of $\widetilde{\bB}$ (or equivalently, a row permutation of $\bW$) that is \textit{admissible}. Specifically, a row permutation of $\bW$ is admissible if the resulting matrix has all non-zero diagonal elements. This property arises from the assumption that the diagonal elements of $\bB$ are all zero to avoid self-loops, which in turn ensures that the diagonal entries of $\bW=\bI-\bB$ are non-zero (up to a scaling factor). By normalizing the rows of $\bW$ such that all diagonal elements equal one, this unique admissible row permutation of $\bW$ defines a unique representation of $\bB$, and consequently, a unique graphical representation of the DAG $\mG$. 

When disjoint cycles are allowed in the underlying true graph $\mG$, multiple admissible row permutations of $\bW$ may exist. However, Theorem 6 in \cite{lacerda2008discovering_supp} shows that among all admissible row permutations of $\bW$, there is a unique row permutation that results in a stable $\bB$. Consequently, this implies the existence of a unique graphical representation of the directed graph $\mG$ with disjoint cycles, which completes the proof.
\end{proof}

\subsection*{A2. Proofs of Non-Closure Under Gaussian Convolution}

\begin{proof}
We prove that the Uniform, Cauchy, Logistic, and Student's $t$-distributions are not closed under convolution with Gaussian distributions.
\paragraph{Uniform distribution} Let $X\sim\operatorname{Unif}[-a,a]$ with $a>0$; its characteristic function is
\begin{equation*}
    \phi_{X}(t)=\frac{\sin(at)}{at}.
\end{equation*}
Let $Y\sim N(0,\sigma^2)$. Assume that $Z=X+Y$ is again uniform on $[-a',a']$ with $a'>0$.  Then
\begin{equation*}
    \phi_{Z}(t)=e^{-\sigma^{2}t^{2}/2}\,\frac{\sin(at)}{at}
    =\frac{\sin(a't)}{a't},
\end{equation*}
for $t\neq \frac{k\pi}{a}$ it rearranges to
\begin{equation*}
    e^{-\sigma^{2}t^{2}/2} \;=\; \frac{at\,\sin(a't)}{a't\,\sin(at)}.
\end{equation*}
Set $t=\frac{\pi}{a'}$, then $\sin(a't)=\sin(\pi)=0$, so the numerator of the right–hand side vanishes. Since $\text{Var}(X) < \text{Var}(Z)$, we have $a<a'$ and the denominator $\sin(at)=\sin(\frac{a\pi}{a'})\neq0$. Hence
\begin{equation*}
    e^{-\sigma^{2}t^{2}/2}=0,
\end{equation*}
which is impossible whenever $\sigma^{2}>0$ because the exponential is strictly positive for all real $t_k$.

\paragraph{Cauchy distribution} Let $X$ be Cauchy with scale parameter $\gamma$ (and location $0$), we have
\begin{equation*}
    \phi_X(t)= e^{-\gamma|t|}.
\end{equation*}
Let $Y\sim N(0,\sigma^2)$. Assuming $Z = X+Y$ is Cauchy with scale parameter $\gamma'$, so that 
\begin{equation*}
    \phi_Z(t)= e^{-\gamma|t|}\, e^{-\sigma^2t^2/2} = e^{-\gamma'|t|},
\end{equation*}
which implies
\begin{equation*}
    \sigma^2t^2/2 = (\gamma' - \gamma)|t|.
\end{equation*}
This reduces to a root finding problem for quadratic equation in $|t|$, and the number of solution does not exceed two. We conclude that it will not hold for all $t$.

\paragraph{Logistic distribution} let $X\sim\text{Logistic}(0,s)$.  Its second and fourth cumulants are
\begin{equation*}
\kappa_2(X)\;=\;\text{Var}(X)\;=\;\frac{\pi^2s^2}{3},
\quad
\kappa_4(X)\;=\;\frac{2\pi^4s^4}{15},
\end{equation*}
Let $Y\sim N(0,\sigma^2)$, so
\begin{equation*}
\kappa_2(Y)=\sigma^2,
\quad
\kappa_k(Y)=0\quad(k\ge3).
\end{equation*}
Since cumulants add under convolution,
\begin{equation*}
    Z = X+Y
\quad\Longrightarrow\quad
\begin{cases}
\kappa_2(Z)
=\kappa_2(X)+\kappa_2(Y)
=\dfrac{\pi^2s^2}{3}+\sigma^2,\\[1em]
\kappa_4(Z)
=\kappa_4(X)+\kappa_4(Y)
=\dfrac{2\pi^4s^4}{15}.
\end{cases}
\end{equation*}
On the other hand, if $Z\sim\text{Logistic}(0,s')$ then
\begin{equation*}
\kappa_2(Z)=\frac{\pi^2{s'}^2}{3},
\quad
\kappa_4(Z)=\frac{2\pi^4{s'}^4}{15}.
\end{equation*}
Equating gives the system
\begin{align*}
\frac{\pi^2{s'}^2}{3}
&=\frac{\pi^2s^2}{3}+\sigma^2,\\
\frac{2\pi^4{s'}^4}{15}
&=\frac{2\pi^4s^4}{15}.
\end{align*}
From the second equation ${s'}^4=s^4$, and since $s',s>0$ we get $s'=s$.  Substituting back yields
\begin{equation*}
\frac{\pi^2s^2}{3}
=\frac{\pi^2s^2}{3}+\sigma^2
\quad\Longrightarrow\quad
\sigma^2=0.
\end{equation*}
Thus, the only solution is $\sigma=0$ and $s'=s$, and no non‐degenerate Gaussian convolution of a logistic random variable can again be logistic.

\paragraph{Student's $t$-distribution}
Write the scale‑mixture representation of Student's $t$ random variable $X\mid W\sim N\!(0,\nu/W)$ with $W\sim\chi^{2}_{\nu}$.
For $Y\sim N(0,\sigma^{2})$ independent of $(X,W)$,
\begin{equation*}
    Z:=X+Y\;\Big|\;W\;\sim\;N\!(0,\;\sigma^{2}+\tfrac{\nu}{W})
          =N(0,V), \text{ where } V:=\sigma^{2}+\tfrac{\nu}{W}.
\end{equation*}
Thus $V\ge\sigma^{2}>0$ almost surely. Any Student's $t$ random variable $T\sim t_{\nu'}$ satisfies $T\mid W'\sim N\!(0,\nu'/W')$ with $W'\sim\chi^{2}_{\nu'}$, so its mixing variance $V'=\nu'/W'$ has support $(0,\infty)$. Since the supports of $V$ and $V'$ differ unless $\sigma^{2}=0$, $Z$ cannot be Student's $t$ for any $\nu'$.
\end{proof}

\subsection*{A3. Proof of Lemma~2}
\begin{proof}
We first show that the independent non-Gaussian error variance $\bSigma_E$ is identifiable from the observed covariance $\bOmega = \bL\bL^{\mT} + \bSigma_E$ under the Assumption~2 (a), and then demonstrate that the latent confounding effects $\bL$ are uniquely identifiable under the Assumption~2 (b).

Consider the FA form of model $\widetilde{\bY} = \bL\bC + \bE$  in the manuscript, and let $\bSigma_{L}=\bL\bL^{\mT}$. Suppose that the non-Gaussian error $E_q$, $q=1,2,\dots,Q$ satisfies Assumption~2 (a). Assume that there exists another set of factor loading matrix $\bL^*$ and exogenous error $\bE^*$ satisfying $\widetilde{\bY} = \bL^*\bC + \bE^*$, where $E^*_q$ follows the same non-Gaussian distribution form as $E_q$ for any $q$. Then $\bOmega = \bL^*(\bL^*)^{\mT} + \bSigma_{E^*} = \bSigma_{L^*} + \bSigma_{E^*} = \bSigma_{L} + \bSigma_{E}$. Assume that at least one diagonal element of $\bSigma_{L^*}$ is different from $\bSigma_{L}$; otherwise, $\bSigma_{E^*}$ and $\bSigma_{E}$ are identical immediately. 

Now we rewrite $\widetilde{\bY} = \bL\bC + \bE$ and $\widetilde{\bY} = \bL^*\bC + \bE^*$ as $\widetilde{\bY} = \widetilde{\bC} + \bE$ and $\widetilde{\bY} = \widetilde{\bC}^* + \bE^*$, where $\widetilde{\bC}$ and $\widetilde{\bC}^*$ are two zero-mean multivariate Gaussian random variables defined by the covariance matrices $\bSigma_{L}$ and $\bSigma_{L^*}$, respectively. Define the variance gap between $\widetilde{C}_q$ and $\widetilde{C}_q^*$ as $(\sigma_{\Delta q})^{2}=|(\bSigma_{L^{*}})_{qq}- (\bSigma_{L})_{qq}|$, $q=1,\dots,Q$, and introduce the following auxiliary variables:
\begin{equation*}
\begin{cases}
c_{\Delta q}=0, & \text{if } (\sigma_{\Delta q})^{2}=0,\\
c_{\Delta q} \sim \mathcal{N}(0,(\sigma_{\Delta q})^{2}), & \text{if } (\sigma_{\Delta q})^{2}>0,
\end{cases}
\quad q=1,\dots,Q.
\end{equation*}
We further define two $Q$-dimensional vectors $\bC_{\Delta}$ and $\bC_{\Delta^*}$, where $(\bC_{\Delta})_q = c_{\Delta q} \cdot \mathbb{I}[(\bSigma_{L})_{qq}\le (\bSigma_{L^{*}})_{qq}]$ and $(\bC_{\Delta^*})_q= c_{\Delta q} \cdot \mathbb{I}[(\bSigma_{L^{*}})_{qq}\le (\bSigma_{L})_{qq}]$, for $q=1,\dots,Q$. Note that for any $q$, if $(\bSigma_{L^{*}})_{qq} = (\bSigma_{L})_{qq}$, then $(\sigma_{\Delta q})^2 = 0$, which implies $c_{\Delta q} = 0$. In this case, $(\bC_{\Delta})_q= (\bC_{\Delta^*})_q = 0$. In contrast, if $(\bSigma_{L})_{qq}$ and $(\bSigma_{L^{*}})_{qq}$ differ, then $c_{\Delta q}$ is assigned to the vector corresponding to the smaller variance, while the other vector receives a zero at that position.

By introducing the auxiliary variables, we have 
\begin{equation*}
\widetilde{\bY} = \widetilde{\bC} + \bC_{\Delta} + \bE - \bC_{\Delta} = \widetilde{\bC}^* + \bC_{\Delta^*} + \bE^* - \bC_{\Delta^*},
\end{equation*}
where $\widetilde{\bC} + \bC_{\Delta}$ and $\widetilde{\bC}^* + \bC_{\Delta^*}$ are identical multivariate Gaussian distributions by construction. 
Therefore, $\bE - \bC_{\Delta}$ and $\bE^* - \bC_{\Delta^*}$ have the same distribution form, or equivalently, $E_q - (\bC_{\Delta})_q$ and $E^*_q - (\bC_{\Delta^*})_q$ are in the same distribution form, for any $q=1,\dots,Q$. Without loss of generality, assume that $(\bSigma_{L})_{qq}\le (\bSigma_{L^{*}})_{qq}$ for any $q$, then $E_q - c_{\Delta q}$ and $E^*_q$ have the same distribution form. However, this means that the convolution of the $q$-th independent non-Gaussian error term $E_q$ and a Gaussian random variable $c_{\Delta q}$ results in another random variable $E^*_q$ from the same non-Gaussian distribution, which contradicts with the Assumption~2 (a) in the manuscript. Therefore, we conclude that $\bSigma_E$ is uniquely identifiable.

Next, we will show that the latent confounding effects $\bL$ are uniquely identifiable. Note that the unique identification of $\bSigma_E$ implies that $\bL\bL^T$ is uniquely identified. We then follow the idea of \cite{fruhwirth2023counts_supp} to show that $\bL$ is identifiable up to column switching and sign flips under Assumption~2 (b) in the manuscript. 

Define the ordered generalized lower triangular (GLT) structure based on the unordered GLT (UGLT) structure, with an additional constraint that the distinct pivot rows $\ell_1,\dots,\ell_P$ are ordered, i.e., $\ell_1 < \dots < \ell_P$. Let $\bL_{ord}$ denote the ordered GLT matrix by applying column permutation matrix $\bm{P}_{per}$ and the sign flip matrix $\bm{P}_{sign}$ to $\bL$, i.e., $\bL_{ord} = \bL \bm{P}_{per}\bm{P}_{sign}$. Consider another factor loading matrix with unordered GLT structure $\widetilde{\bL}$ that satisfies $\bL\bL^T = \widetilde{\bL}\widetilde{\bL}^T$. Similarly, let $\widetilde{\bL}_{ord}$ denote the GLT matrix obtained by applying column permutation matrix $\bm{\widetilde{P}}_{per}$ and sign flip matrix $\bm{\widetilde{P}}_{sign}$ to $\widetilde{\bL}$, i.e., $\widetilde{\bL}_{ord} = \widetilde{\bL} \bm{\widetilde{P}}_{per}\bm{\widetilde{P}}_{sign}$. By Corollary~1 from \cite{fruhwirth2023counts_supp}, we have $\bL_{ord} = \widetilde{\bL}_{ord}$, which implies $\bL \bm{P}_{per}\bm{P}_{sign}=\widetilde{\bL} \bm{\widetilde{P}}_{per}\bm{\widetilde{P}}_{sign}$, i.e., $\bL$ is identifiable up to column switching and sign flips. 

Lastly, column permutations and sign flips correspond to label switching and sign indeterminacy of the latent confounders $\bC$, and the ordering and sign of the columns carry no meaningful interpretation as they are latent variables. Therefore, without loss of generality, we conclude that $\bL$ is uniquely identified, which completes the proof.
\end{proof}

\subsection*{A4. Proof of Corollary~1}
\begin{proof}
We first show that under the full model
\begin{equation}
    \label{eqn:full_model_supp}
    \bY = \bmu + \bB\bY + \bA\bX + \bL\bC + \bE,
\end{equation}
$\bB$ is still uniquely identifiable given the Assumptions~1 (a)-(b). Rewrite the model as
\begin{equation*}
\bY = (\bI-\bB)^{-1}(\bmu + \bA\bX + \bL\bC + \bE) =
\begin{bmatrix}
    \widetilde{\bB},\widetilde{\bB},\, \widetilde{\bB}\bL
\end{bmatrix}
\begin{bmatrix}
\bmu + \bA\bX \\
\bE \\
\bC
\end{bmatrix}
= \begin{bmatrix}
    \widetilde{\bB}, \widetilde{\bB},\, \widetilde{\bL}
\end{bmatrix}
\begin{bmatrix}
\bmu + \bA\bX \\
\bE \\
\bC
\end{bmatrix},
\end{equation*}
where $\widetilde{\bB} = (\bI-\bB)^{-1}$ and $\widetilde{\bL} = \widetilde{\bB}\bL$. Then the same argument from the proof of Lemma~1 proves that $\bB$ is uniquely identified. We can then rewrite the full model~\ref{eqn:full_model_supp} as $\widetilde{\bY}=\bL\bC + \bE$, where $\widetilde{\bY}=\widetilde{\bB}^{-1}\bY-(\bmu + \bA\bX)$. Note that $\bmu + \bA\bX$ can be identified from the mean value of the identified components $\widetilde{\bB}^{-1}\bY$, and the covariance matrix of $\widetilde{\bY}$ can still be decomposed as $\bOmega = \bL\bL^{\mT} + \bSigma_E$. Consequently, the unique identification of $\bL$ follows from the same argument as the proof of Lemma~2 with Assumptions~2 (a)-(b), which completes the proof.

\end{proof}

\clearpage
\section*{B. Details of RJMCMC Algorithm}

\subsection*{B1. Update \texorpdfstring{$\bmu$}{mu}}
The full conditional for $\mu_q$ is proportional to:
\begin{equation*}
  p(\mu_q \mid \cdot)
  \propto p(\mu_q) \prod_{i=1}^n \mathcal{N} (Y_{iq} \mid \widetilde{Y}_{iq}, \sigma_q^2/\tau_{iq}),
\end{equation*}
where $\widetilde{Y}_{iq} = (\bmu + \bA\bX_i + \bB\bY_i + \bL\bC_i)_q$.
Then $\mu_q$ can be updated via:
\begin{equation*}
  \mu_q \mid \cdot \sim \mathcal{N}(\mu_{n},\,V_{n}),
\end{equation*}
where 
\begin{equation*}
  V_{n} = \left(\frac{1}{\sigma_{\mu}^{2}} + \sum_{i=1}^n \frac{\tau_{iq}}{\sigma_{q}^{2}}\right)^{-1}, \quad
  \mu_{n} = V_{n}\,\sum_{i=1}^n \tfrac{\tau_{iq}}{\sigma_{q}^{2}}\widetilde{E}_{iq},
\end{equation*}
and $\widetilde{E}_{iq} = (\bY_i - \bA\bX_i - \bB\bY_i - \bL\bC_i)_q$.

\subsection*{B2. Update \texorpdfstring{$\bA$}{A} and related parameters}

\subsubsection*{B2.1. Update \texorpdfstring{$\bA$}{A}}
The full conditional for the $q$-th row of $\bA$, denoted as $\bA_{q\cdot}$, is
\begin{equation*}
  p(\bA_{q\cdot}\mid\cdot) \propto p(\bA_{q\cdot}) \prod_{i=1}^{n} \mathcal{N}(Y_{iq}\mid \widetilde{Y}_{iq}, \sigma_{q}^{2}/\tau_{iq}),
\end{equation*}
where $\widetilde{Y}_{iq} = (\bmu + \bA\bX_i + \bB\bY_i + \bL\bC_i)_q$.
Then $\bA_{q,\cdot}$ can be updated via: 
\begin{equation*}
  \bA_{q\cdot}\mid\cdot \sim \mathcal{N}(\bmu_{n},\,\bV_{n}),
\end{equation*}
where
\begin{equation*}
  \bV_{n} = \left(\sum_{i=1}^{n} \frac{\tau_{iq}}{\sigma_{q}^{2}}\bX_i\bX_i^{\mT} + \text{diag}\left(\tfrac{1}{\gamma_{qs}^{\alpha}\nu_{qs}^{\alpha}}\right)_{s=1,\ldots,S}\right)^{-1}, \quad
  \bmu_{n} = \bV_{n} \sum_{i=1}^{n}\frac{\tau_{iq}}{\sigma_{q}^{2}}\bX_i\widetilde{E}_{iq},
\end{equation*}
and $\widetilde{E}_{iq} = (\bY_i - \bmu - \bB\bY_i - \bL\bC_i)_q$.

\subsubsection*{B2.2. Update \texorpdfstring{$\gamma^\alpha_{qs}$}{gamma\^alpha\_qs}}

The full conditional of $\gamma^\alpha_{qs}$ can be written as
\begin{equation*}
  p(\gamma^\alpha_{qs}\mid\cdot) \propto p(\gamma^\alpha_{qs}) p(A_{qs}).
\end{equation*}
We obtain the following odds to update $\gamma^\alpha_{qs}$:
\begin{equation*}
  \frac{p(\gamma^\alpha_{qs}=1\mid\cdot)}{p(\gamma^\alpha_{qs}=\nu_0\mid\cdot)} = \frac{\sqrt{\nu_0}\rho_\alpha}{1-\rho_\alpha}
  \exp\left(\frac{(1-\nu_0)A_{qs}^2}{2\nu_0\nu^\alpha_{qs}}\right).
\end{equation*}

\subsubsection*{B2.3. Update \texorpdfstring{$\nu^\alpha_{qs}$}{nu\_alpha\_qs}}

The full conditional of $\nu^\alpha_{qs}$ can be written as
\begin{equation*}
    p(\nu^\alpha_{qs}\mid\cdot) \propto p(\nu^\alpha_{qs})p(A_{qs}).
\end{equation*}
We obtain the following update for $\nu^\alpha_{qs}$:
\begin{equation*}
  \nu^\alpha_{qs} \sim \text{Inverse-Gamma}\left(a_\nu + \tfrac12,\, b_\nu + \tfrac{A_{qs}^2}{2\gamma^\alpha_{qs}}\right).
\end{equation*}

\subsubsection*{B2.4. Update \texorpdfstring{$\rho_{\alpha}$}{rho\_alpha}}

We have the following full conditional for $\rho_{\alpha}$:
\begin{equation*}
  p(\rho_\alpha \mid \cdot) \propto p(\rho_\alpha) \prod_{q=1}^{Q} \prod_{s=1}^{S} p(\gamma^\alpha_{qs}),
\end{equation*}
thus we can update $\rho_{\alpha}$ using:
\begin{equation*}
  \rho_{\alpha} \sim \text{Beta}\left(a_\rho + \sum_{q=1}^Q \sum_{s=1}^S \mathbb{I}\{\gamma^\alpha_{qs}=1\}, 
  b_\rho + \sum_{q=1}^Q \sum_{s=1}^S \mathbb{I}\{\gamma^\alpha_{qs}=\nu_0\}\right).
\end{equation*}

\subsection*{B3. Update \texorpdfstring{$\bB$}{B} and related parameters}

The full conditional for $B_{qq'}$ can be written as
\begin{equation*}
  p(B_{qq'}\mid\cdot) \propto p(B_{qq'}) \prod_{i=1}^{n} \mathcal{N}(Y_{iq}\mid\widetilde{Y}_{iq}, \sigma_{q}^{2}/\tau_{iq}) \lvert \bI - \bB \rvert,
\end{equation*}
where $\widetilde{Y}_{iq} = (\bmu + \bA\bX_i + \bB\bY_i + \bL\bC_i)_q$, and the additional term $\lvert \bI - \bB \rvert$ is the Jacobian matrix due to change-of-variable. Since the full conditional distribution for $B_{qq'}$ depends on $\lvert \bI - \bB \rvert$, there is no closed‑form solution. Hence, we update $B_{qq'}$ using the Metropolis-Hastings (MH) algorithm. At each MH iteration, we accept the move only if the largest eigenvalue modulus of $\bB$ is strictly less than $1$, ensuring the proposed model is stable and well‑defined. The rest of the parameters $\gamma^\beta_{qq'}$, $\nu^\beta_{qq'}$, and $\rho_{\beta}$ can be updated in the same way as in the update of $\bA$.

\subsection*{B4. Update \texorpdfstring{$\bL$}{L} and related parameters}

\subsubsection*{B4.1. Update \texorpdfstring{$\bL$}{L}}
Let $\bL_{q,K_q}$ denote the vector of non‑zero entries of the $q$‑th row of $\bL$, where $K_q = \{\, p \in \{1,2,\dots,P^*\} : \delta_{qp} \neq 0 \}$. If all the entries are zero in the $q$‑th row, then skip to update the next row; otherwise the full conditional for $\bL_{q,K_q}$ is
\begin{equation*}
  p(\bL_{q,K_q}\mid\cdot) \propto p(\bL_{q,K_q}) \prod_{i=1}^{n} \mathcal{N}(Y_{iq}\mid\widetilde{Y}_{iq}, \sigma_{q}^{2}/\tau_{iq}),
\end{equation*}
where $\widetilde{Y}_{iq} = (\bmu + \bA\bX_i + \bB\bY_i + \bL\bC_i)_q$. Thus we can update $\bL_{q,K_q}$ via
\begin{equation*}
  \bL_{q,K_q}\mid\cdot \sim \mathcal{N}(\bmu_{n}, \sigma_{q}^{2}\bV_{n}),
\end{equation*}
where
\begin{equation*}
  \bV_{n} = \left( \sum_{i=1}^{n} \tau_{iq}\,\bC_{i,K_q}\bC_{i,K_q}^{\mT} + \text{diag}\bigl(\tfrac{1}{\kappa}\bigr)_{|K_q|}\right)^{-1},
  \quad \bmu_{n} = \bV_{n} \sum_{i=1}^{n} \tau_{iq}\,\bC_{i,K_q}\,\widetilde{E}_{iq},
\end{equation*}
and $\widetilde{E}_{iq} = (\bY_i - \bmu - \bA\bX_i - \bB\bY_i)_q$.

\subsubsection*{B4.2. Update \texorpdfstring{$\bdelta$}{delta}}

For non‑zero column $p$, update each $\delta_{qp}$ where $q>\ell_p$, i.e., the rows below the pivot row $\ell_p$. The full conditional for $\delta_{qp}$ is
\begin{equation*}
  p(\delta_{qp}\mid\cdot) \propto p(\delta_{qp}) \prod_{i=1}^{n} \mathcal{N}(Y_{iq}\mid\widetilde{Y}_{iq},\sigma_{q}^{2}/\tau_{iq}),
\end{equation*}
where $\widetilde{Y}_{iq} = (\bmu + \bA\bX_i + \bB\bY_i + \bL\bC_i)_q$. The full‑conditional odds for $\delta_{qp}$  is:
\begin{equation*}
  \frac{p(\delta_{qp}=1\mid\cdot)}{p(\delta_{qp}=0\mid\cdot)} =
  \frac{\zeta_p\prod_{i=1}^{n}\mathcal{N}(Y_{iq}\mid\widetilde{Y}_{iq},\sigma_{q}^{2}/\tau_{iq},\delta_{qp}=1)}
       {(1-\zeta_p)\prod_{i=1}^{n}\mathcal{N}(Y_{iq}\mid\widetilde{Y}_{iq},\sigma_{q}^{2}/\tau_{iq},\delta_{qp}=0)}.
\end{equation*}

\subsubsection*{B4.3. Update \texorpdfstring{$\bm{\ell}$}{ell}}
We update each pivot in the pivot rows of $\bm{\ell}$ with an MH step following  \cite{fruhwirth2024sparse_supp}. At each MH iteration, one of the three moves, shifting, switching, adding/deleting, is proposed with the probability $p_{shift}$, $p_{switch}$, $1 - p_{shift} - p_{switch}$, respectively. If adding/deleting is proposed, there is a probability of $p_{add}$ to add a pivot, and $1 - p_{add}$ to delete a pivot. We describe the details of these three moves below.

\paragraph{Shifting the pivot.}
Let $\ell_{\star}$ denote the index of the first nonzero row below $\ell_{p}$ (i.e., $\delta_{\ell_{\star},p} = 1$, $\delta_{qp} = 0$ for $\ell_p < q < \ell_{\star}$). If $\ell_{\star} > 2$, then a new pivot $\ell_{p}^{\text{new}}$ is sampled uniformly at random from the set $\mathcal{M}_p=\{1,\ldots, \ell_{\star}-1\}\cap\mathcal{Q}(\bm\ell)$. This set $\mathcal{M}_p$ consists of row indices that have not been used as pivots. If $\mathcal{M}_p$ is empty, then no shift move is performed.

\paragraph{Switching pivots.}
This move is only performed if $P^*>1$. We randomly select a nonzero column $p'\neq p$. For all rows $q$ between (and including) $\min(\ell_{p'}, \ell_{p})$ and $\max(\ell_{p'}, \ell_{p})$ in which the indicators $\delta_{qp}$ and $\delta_{qp'}$ differ, we swap the indicators between the two columns. Formally, $\delta_{qp}^{\text{new}}=1 - \delta_{qp}$, $\delta_{qp'}^{\text{new}}=1 - \delta_{qp'}$, for all $q\in
  \{\,\min(\ell_{p'},\ell_{p}),\dots,\max(\ell_{p'},\ell_{p})\}$ with $\delta_{qp}\neq \delta_{qp'}$.

\paragraph{Adding or deleting a pivot.}
For the adding move, we introduce a new pivot $\ell_{p}^{\text{new}}$ above the current pivot $\ell_{p}$, in a row that is not already a pivot in any other column. Specifically, we sample $\ell_{p}^{\text{new}}$ uniformly from the set $\mathcal{A}_p\;=\; \{\,1,\dots, \ell_{p}-1\}\cap\mathcal{Q}(\bm\ell)$, and set $\delta_{\ell_{p}^{\text{new}},\,p} = 1$, while leaving all other elements of $\bdelta$ (including $\delta_{\ell_{p},\,p}=1$) unchanged. Deleting move is the reverse of the adding move that removes the current pivot $\ell_{p}$ and makes the first nonzero row below $\ell_{p}$ (denoted $\ell_{\star}$) the new pivot, i.e., $\delta_{\ell_{p},p}^{\text{new}}=0$. All other entries remain unchanged. We only delete a pivot if $\ell_{\star}$ is not used as a pivot in any other column, i.e., $\ell_{\star}\in\mathcal{Q}(\bm\ell)$. \\

Lastly, if one of the moves is proposed in an MCMC iteration, we run the standard MH algorithm to accept or reject the proposal.

\subsubsection*{B4.4. Update \texorpdfstring{$\kappa$}{kappa}}

Since $\kappa$ has the full conditional:
\begin{equation*}
 p(\kappa\mid \cdot) \propto p(\kappa) \prod_{q,p} p(L_{qp}),
\end{equation*}
we can update it via
\begin{equation*}
  \kappa \sim \mathrm{Inverse\text{-}Gamma}(a_{\text{new}},\,b_{\text{new}}),
\end{equation*}
where
\begin{equation*}
  a_{\text{new}} = a_\kappa + \tfrac12\sum_{q,p}\delta_{qp}, \text{ and }
  b_{\text{new}} = b_\kappa + \tfrac12\sum_{q,p}\frac{\delta_{qp}L_{qp}^{2}}{\sigma_{q}^{2}}.
\end{equation*}

\subsubsection*{B4.5. Update \texorpdfstring{$\zeta_p$}{zeta\_p}}
Since $\zeta_p$ has the full conditional:
\begin{equation*}
  p(\zeta_p\mid\cdot) \propto p(\zeta_p) \prod_{q} p(L_{qp}),
\end{equation*}
we can update it via:
\begin{equation*}
  \zeta_p \sim  \text{Beta}(\alpha_p,\beta_p),
\end{equation*}
where
\begin{equation*}
  \alpha_p = \frac{a_1 a_2}{P} + d_p - 1, \quad
  \beta_p = a_2 + Q - \ell_{p} - d_p + 1, 
\end{equation*}
and $d_p = \sum_{q=1}^{Q}\delta_{qp}$.

\subsubsection*{B4.6. Update \texorpdfstring{$a_1, a_2$}{a1, a2}}

The full conditional distribution for $a_1, a_2$ is
\begin{equation*}
    p(a_1, a_2 \mid \cdot) \;\propto\; p(a_1,a_2) \prod_{p=1}^{P^*} p(\zeta_p).
\end{equation*}
There is no closed form for the full conditional of $a_1$ and $a_2$, thus we update them via standard MH steps.

\subsection*{B5. Update \texorpdfstring{$\bC$}{C}}
The full conditional distribution for $\bC_i$ can be written as
\begin{equation*}
  p(\bC_i \mid \cdot) \propto p(\bC_i) \prod_{q=1}^{Q} \mathcal{N}(Y_{iq} \mid \widetilde{Y}_{iq}, \sigma_{q}^{2}/\tau_{iq}),
\end{equation*}
where $\widetilde{Y}_{iq} = (\bmu + \bA\bX_i + \bB\bY_i + \bL\bC_i)_q$. Thus we can update $\bC_i$ via:
\begin{equation*}
  \bC_i \sim \mathcal{N}(\bmu_n, \bV_n),
\end{equation*}
where
\begin{equation*}
  \bV_n = \left(\bI_{P^*} + \bL^{\mT}\,\bSigma_{i}^{-1}\bL\right)^{-1},
  \quad
  \bmu_n = \bV_n\,\bL^{\mT}\,\bSigma_{i}^{-1}\, \widetilde{\bE}_i,
\end{equation*}
and $\bSigma_{i}^{-1}=\text{diag}\left(\frac{\tau_{iq}}{\sigma_{q}^{2}}\right)_{q=1,\dots,Q}$, $\widetilde{\bE}_{i} = \bY_i - \bmu - \bA\bX_i - \bB\bY_i$.

\subsection*{B6. Update \texorpdfstring{$\bE$}{E} and related parameters}

\subsubsection*{B6.1. Update \texorpdfstring{$\tau$}{tau}}

The full conditional distribution for $\tau_{iq}$ is
\begin{equation*}
  p(\tau_{iq}\mid\cdot) \propto p(\tau_{iq}) \prod_{i=1}^{n} \mathcal{N}(Y_{iq}\mid\widetilde{Y}_{iq}, \sigma_{q}^{2}/\tau_{iq}).
\end{equation*}
$\tau_{iq}$ can then be updated via:
\begin{equation*}
  \tau_{iq}\mid\cdot \sim \mathrm{Inverse\text{-}Gaussian}\left( \frac{\sigma_{q}}{2\lvert Y_{iq}-\widetilde{Y}_{iq}\rvert}, \frac14\right).
\end{equation*}

\subsubsection*{B6.2. Update \texorpdfstring{$\sigma_q^2$}{sigma2q}}

The full conditional distribution for $\sigma_q^2$ is
\begin{equation*}
  p(\sigma_q^{2}\mid\cdot) \propto p(\sigma_q^{2}) \prod_{i=1}^{n} \mathcal{N}(Y_{iq}\mid\widetilde{Y}_{iq},\sigma_{q}^{2}/\tau_{iq}).
\end{equation*}
Therefore, $\sigma_q^2$ can be updated via
\begin{equation*}
  \sigma_q^{2}\mid\cdot \sim
  \mathrm{Inverse\text{-}Gamma}\Bigl(a_\sigma + \tfrac{n}{2}, b_\sigma + \tfrac12\sum_{i=1}^{n}(Y_{iq}-\widetilde{Y}_{iq})^{2}\tau_{iq}\Bigr).
\end{equation*}

\subsection*{B7. Update \texorpdfstring{P\textsuperscript{*}}{P*}}

In this section, we outline the computation details of acceptance probability $MH_{split}(P^*, P_{single})$. For split move, the acceptance probability can be computed analogously and satisfies $MH_{merge}(P^*, P_{single}) = (MH_{split}(P^* - 1, P_{single} - 1))^{-1}$.
When proposing a split move on one of the $P-P^*$ zero columns $p$ in $\bdelta$, we first sample a pivot row $\ell_p$ for column $p$ from the set of pivot positions not yet assigned to the existing $P^*$ non-zero columns. We then define the following transformation between different model spaces, mapping from $(\sigma_{\ell_p}^2, U)$ to $((\sigma_{\ell_p}^2)^{new}, L_{\ell_{p},p}^{new})$, specified as $(\sigma_{\ell_p}^2)^{new} = (1 - U^2)\sigma_{\ell_p}^2$ and $L_{\ell_{p},p}^{new} = \sqrt{8\sigma_{\ell_p}^2}U$, where $U$ is an auxiliary variable sampled from the uniform distribution on the interval $(0,1)$. The acceptance probability for the split move is given by $\min(1,MH_{split}(P^*,P_{single}))$, where 
\begin{gather*}
    MH_{split}(P^*,P_{single}) = \textit{prior ratio} \times \textit{likelihood ratio} \times \textit{proposal ratio}^{-1} \times \lvert \textit{Jacobian} \rvert \nonumber \\
    = \frac{p(L_{\ell_{p},p}^{new}\mid (\sigma_{\ell_p}^2)^{new}, \delta_{\ell_p,p}^{new})p((\sigma_{\ell_p}^2)^{new})p(\delta_{\ell_p,p}^{new})}{p(L_{\ell_{p},p} \mid \sigma_{\ell_p}^2, \delta_{\ell_p,p})p(\sigma_{\ell_p}^2)p(\delta_{\ell_p,p})} \times 
    \frac{p(Y_{\ell_p}\mid \bL_{\ell_p}^{new}, (\sigma_{\ell_p}^2)^{new})}{p(Y_{\ell_p}\mid \bL_{\ell_p}, \sigma_{\ell_p}^2)} \times
    \frac{P-P^*}{P_{single}+1} \times \sqrt{8\sigma_{\ell_p}^2}.
\end{gather*}
To be more specific, define $a_P = \frac{a_1a_2}{H}$, and the prior ratio can be calculated via:
\begin{equation*}
  \frac{\mathcal{N}(L_{\ell_{p},p}^{new};0,\kappa(\sigma_{\ell_p}^2)^{new}), \delta_{\ell_p,p}^{new})IG((\sigma_{\ell_p}^2)^{new};a_{\sigma},b_{\sigma})a_P(P-P^*)}{IG(\sigma_{\ell_p}^2;a_{\sigma},b_{\sigma})(a_2-1+ P-P^*)}
\end{equation*}
The likelihood ratio can be calculated via:
\begin{equation*}
 \prod_{i=1}^{n} \frac{\mathcal{N}(Y_{i, \ell p} \mid (\widetilde{Y}_{i, \ell p})^{new},(\sigma_{\ell_p}^2)^{new}/\tau_{q,\ell p})}{\mathcal{N}(Y_{i, \ell p} \mid \widetilde{Y}_{i, \ell p},\sigma_{\ell_p}^2/\tau_{q,\ell p})},
\end{equation*}
where $(\widetilde{Y}_{i, \ell p})^{new} = (\bmu + \bA\bX_i + \bB\bY_i + (\bL)^{new}\bC_i)_{\ell p}$ and $\widetilde{Y}_{i, \ell p} = (\bmu + \bA\bX_i + \bB\bY_i + \bL\bC_i)_{\ell p}$.
\clearpage

\section*{C. Supplementary Tables and Figures}

\begin{table*}[htbp]
  \centering
  \renewcommand{\arraystretch}{1.5}
  \begin{tabular}{|c|p{5.5cm}|p{5.5cm}|}
    \hline
    \textbf{Edge Types} & \textbf{Present Relationships} & \textbf{Absent Relationships} \\
    \hline
    A $\rightarrow$ B & A is a cause of B. It may be a direct or indirect cause that may include other measured variables. Also, there may be an unmeasured confounder of A and B. & B is not a cause of A \\
    \hline
    A $\leftrightarrow$ B & There is an unmeasured confounder (call it L) of A and B. There may be measured variables along the causal pathway from L to A or from L to B. & A is not a cause of B. B is not a cause of A. \\
    \hline
    A $\bullet \rightarrow$ B & Either A is a cause of B (i.e., A $\longrightarrow$ B) or there is an unmeasured confounder of A and B (i.e., A $\longleftrightarrow$ B) or both. & B is not a cause of A. \\
    \hline
    A $\bullet- \bullet$ B & Exactly one of the following holds:
    \begin{enumerate}
    \item A is a cause of B
    \item B is a cause of A
    \item there is an unmeasured confounder of A and B
    \item both 1 and 3
    \item both 2 and 3
    \end{enumerate} & \\
    \hline
  \end{tabular}
  \caption{PAG edge types.}
\end{table*}

\clearpage

\begin{table}[!htb]
  \centering
  \small
  \begin{tabular}{lccccc}
    \toprule
        & Truth & Mean (SD) & 95\%CI Coverage & Bias & MSE \\
    \midrule
    $A_{11}$ & $-0.5966$ & $-0.5965\,(0.0105)$ & 1.00 & $1.61{\times}10^{-4}$ & $1.08{\times}10^{-4}$ \\
    $A_{12}$ & $-0.8764$ & $-0.8759\,(0.0090)$ & 1.00 & $5.46{\times}10^{-4}$ & $8.02{\times}10^{-5}$ \\
    $A_{21}$ & $0.7968$  & $0.7905\,(0.0152)$ & 1.00 & $-6.26{\times}10^{-3}$ & $2.66{\times}10^{-4}$ \\
    $A_{22}$ & $-0.5881$ & $-0.5913\,(0.0204)$ & 0.98 & $-3.28{\times}10^{-3}$ & $4.20{\times}10^{-4}$ \\
    $A_{31}$ & $0.8894$  & $0.8829\,(0.0111)$ & 0.98 & $-6.46{\times}10^{-3}$ & $1.63{\times}10^{-4}$ \\
    $A_{32}$ & $-0.6469$ & $-0.6428\,(0.0145)$ & 1.00 & $4.12{\times}10^{-3}$ & $2.22{\times}10^{-4}$ \\
    $A_{41}$ & $0.3216$  & $0.3267\,(0.0183)$ & 0.98 & $5.13{\times}10^{-3}$ & $3.56{\times}10^{-4}$ \\
    $A_{42}$ & $0.3740$  & $0.3741\,(0.0232)$ & 0.98 & $9.59{\times}10^{-5}$ & $5.26{\times}10^{-4}$ \\
    $A_{51}$ & $0.2582$  & $0.2557\,(0.0147)$ & 1.00 & $-2.53{\times}10^{-3}$ & $2.19{\times}10^{-4}$ \\
    $A_{52}$ & $-0.2318$ & $-0.2264\,(0.0160)$ & 0.98 & $5.35{\times}10^{-3}$ & $2.80{\times}10^{-4}$ \\
    \bottomrule
  \end{tabular}
  \caption{Simulation truths, means and standard deviations (SD) of posterior means, 95\% credible interval (CI) coverage probabilities, biases, and mean squared errors (MSE) of the estimated covariate effects of BayCausal under a sample size of 5000 in Simulation Scenario I, based on 50 repeated experiments.}
\end{table}

\begin{table}[!htb]
  \centering
  \small
  \begin{tabular}{lccccc}
    \toprule
        & Truth & Mean (SD) & 95\%CI Coverage & Bias & MSE \\
    \midrule
    $A_{11}$ & $0.8894$  & $0.8886\,(0.0109)$ & 0.98 & $-7.48{\times}10^{-4}$ & $1.17{\times}10^{-4}$ \\
    $A_{12}$ & $-0.2318$ & $-0.2291\,(0.0114)$ & 0.98 & $2.71{\times}10^{-3}$  & $1.34{\times}10^{-4}$ \\
    $A_{21}$ & $0.3216$  & $0.3190\,(0.0226)$ & 0.96 & $-2.58{\times}10^{-3}$ & $5.08{\times}10^{-4}$ \\
    $A_{22}$ & $0.5397$  & $0.5425\,(0.0192)$ & 1.00 & $2.80{\times}10^{-3}$  & $3.70{\times}10^{-4}$ \\
    $A_{31}$ & $0.2582$  & $0.2505\,(0.0199)$ & 1.00 & $-7.76{\times}10^{-3}$ & $4.50{\times}10^{-4}$ \\
    $A_{32}$ & $-0.0046$ & $-0.0001\,(0.0213)$ & 0.96 & $4.46{\times}10^{-3}$  & $4.64{\times}10^{-4}$ \\
    $A_{41}$ & $-0.8764$ & $-0.8772\,(0.0304)$ & 0.94 & $-8.03{\times}10^{-4}$ & $9.04{\times}10^{-4}$ \\
    $A_{42}$ & $0.4352$  & $0.4417\,(0.0212)$ & 0.96 & $6.48{\times}10^{-3}$  & $4.82{\times}10^{-4}$ \\
    $A_{51}$ & $-0.5881$ & $-0.5970\,(0.0234)$ & 0.98 & $-8.95{\times}10^{-3}$ & $6.16{\times}10^{-4}$ \\
    $A_{52}$ & $0.9838$  & $0.9894\,(0.0127)$ & 1.00 & $5.59{\times}10^{-3}$  & $1.89{\times}10^{-4}$ \\
    $A_{61}$ & $-0.6469$ & $-0.6581\,(0.0214)$ & 0.96 & $-1.12{\times}10^{-2}$ & $5.76{\times}10^{-4}$ \\
    $A_{62}$ & $-0.2399$ & $-0.2396\,(0.0146)$ & 1.00 & $3.33{\times}10^{-4}$  & $2.08{\times}10^{-4}$ \\
    $A_{71}$ & $0.3740$  & $0.3717\,(0.0132)$ & 0.98 & $-2.30{\times}10^{-3}$ & $1.77{\times}10^{-4}$ \\
    $A_{72}$ & $0.5549$  & $0.5599\,(0.0142)$ & 0.96 & $4.97{\times}10^{-3}$  & $2.22{\times}10^{-4}$ \\
    \bottomrule
  \end{tabular}
  \caption{Simulation truths, means and standard deviations (SD) of posterior means, 95\% credible interval (CI) coverage probabilities, biases, and mean squared errors (MSE) of the estimated covariate effects of BayCausal under a sample size of 5000 in Simulation Scenario II, based on 50 repeated experiments.}
\end{table}

\begin{table}[!htb]
  \centering
  \small
  \begin{tabular}{llrrrrrrrrr}
    \toprule
    & & \multicolumn{4}{c}{Scenario I} & \multicolumn{4}{c}{Scenario II} \\
    \cmidrule(lr){3-6} \cmidrule(lr){7-10}
    Experiment & Method & CSR & TPR & FDR & MCC & CSR & TPR & FDR & MCC \\
    \midrule
    \multirow{4}{*}{Student's $t$ Errors}
      & BayCausal   & 50 & 1.00 & 0.00 & 1.00 & 50 & 1.00 & 0.00 & 1.00 \\
      & BFCI       & 8 & 0.72 & 0.36 & 0.61 & 0 & 0.63 & 0.34 & 0.58 \\
      & RCD        & 0 & 0.97 & 0.20 & 0.85 & 0 & 0.29 & 0.54 & 0.27 \\
      & \text{LiNG-D} & 7 & 1.00 & 0.46 & 0.67 & 0 & 1.00 & 0.46 & 0.67 \\
      \hline
    \addlinespace[2pt]
    \multirow{4}{*}{Laplace Confounders}
      & BayCausal   & 24 & 1.00 & 0.13 & 0.92 & 45 & 1.00 & 0.02 & 0.99 \\
      & BFCI       & 12 & 0.80 & 0.25 & 0.73 & 0 & 0.64 & 0.27 & 0.62 \\
      & RCD        & 0 & 1.00 & 0.24 & 0.84 & 0 & 0.34 & 0.71 & 0.17 \\
      & \text{LiNG-D} & 0 & 1.00 & 0.37 & 0.75 & 0 & 1.00 & 0.43 & 0.70 \\
    \bottomrule
  \end{tabular}
  \caption{Comparisons of method performance in additional simulation studies, including the counts of successful recovery (CSR), true positive rate (TPR), false discovery rate (FDR), and Matthews correlation coefficient (MCC) across 50 replications.}
\end{table}

\begin{figure}[!htb]
  \centering
  \begin{subfigure}[b]{0.45\textwidth}
      \centering
      \includegraphics[width=\textwidth]{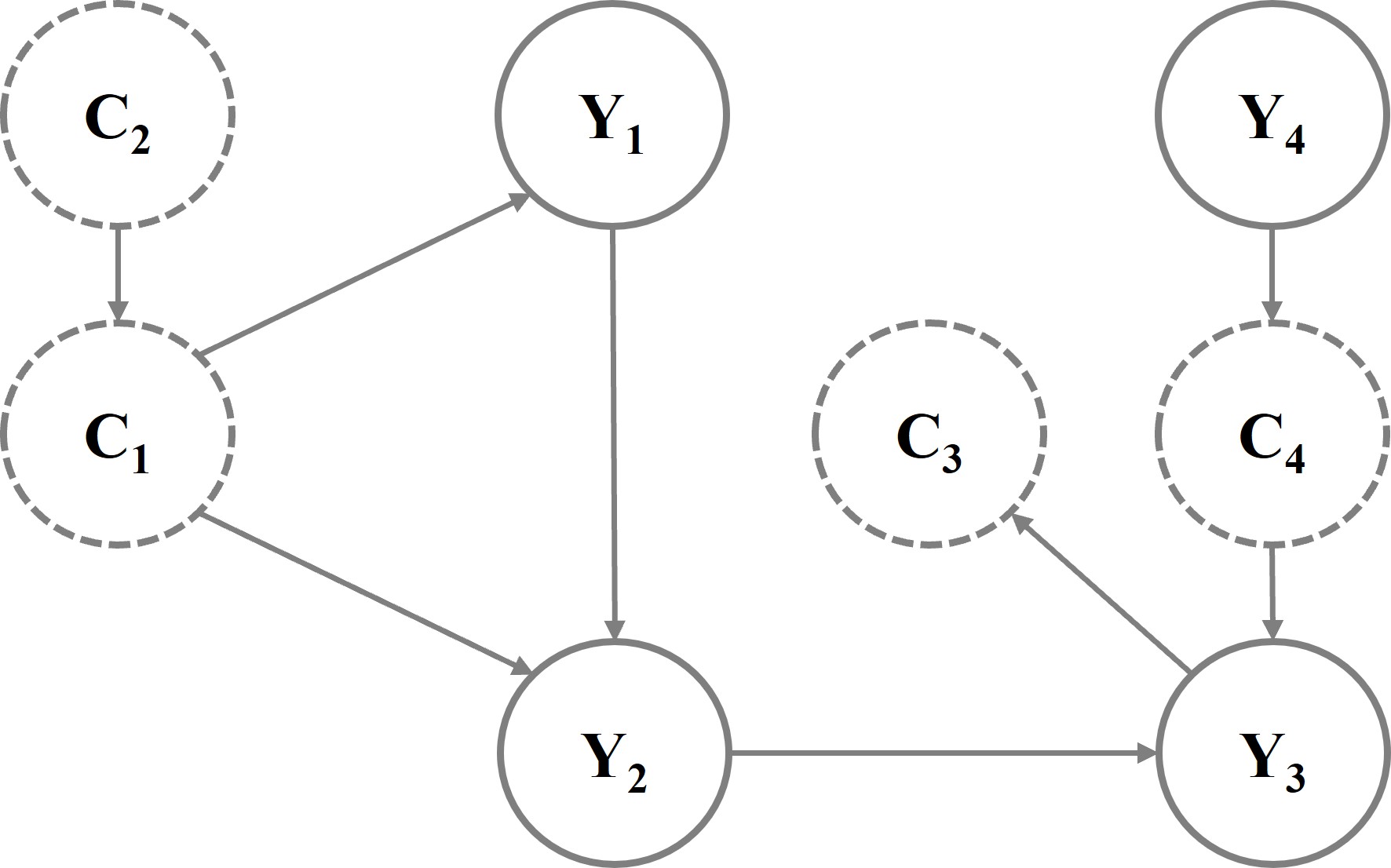}
      (a) Non-canonical form
  \end{subfigure}
  \hfill
  \begin{subfigure}[b]{0.45\textwidth}
      \centering
      \includegraphics[width=\textwidth]{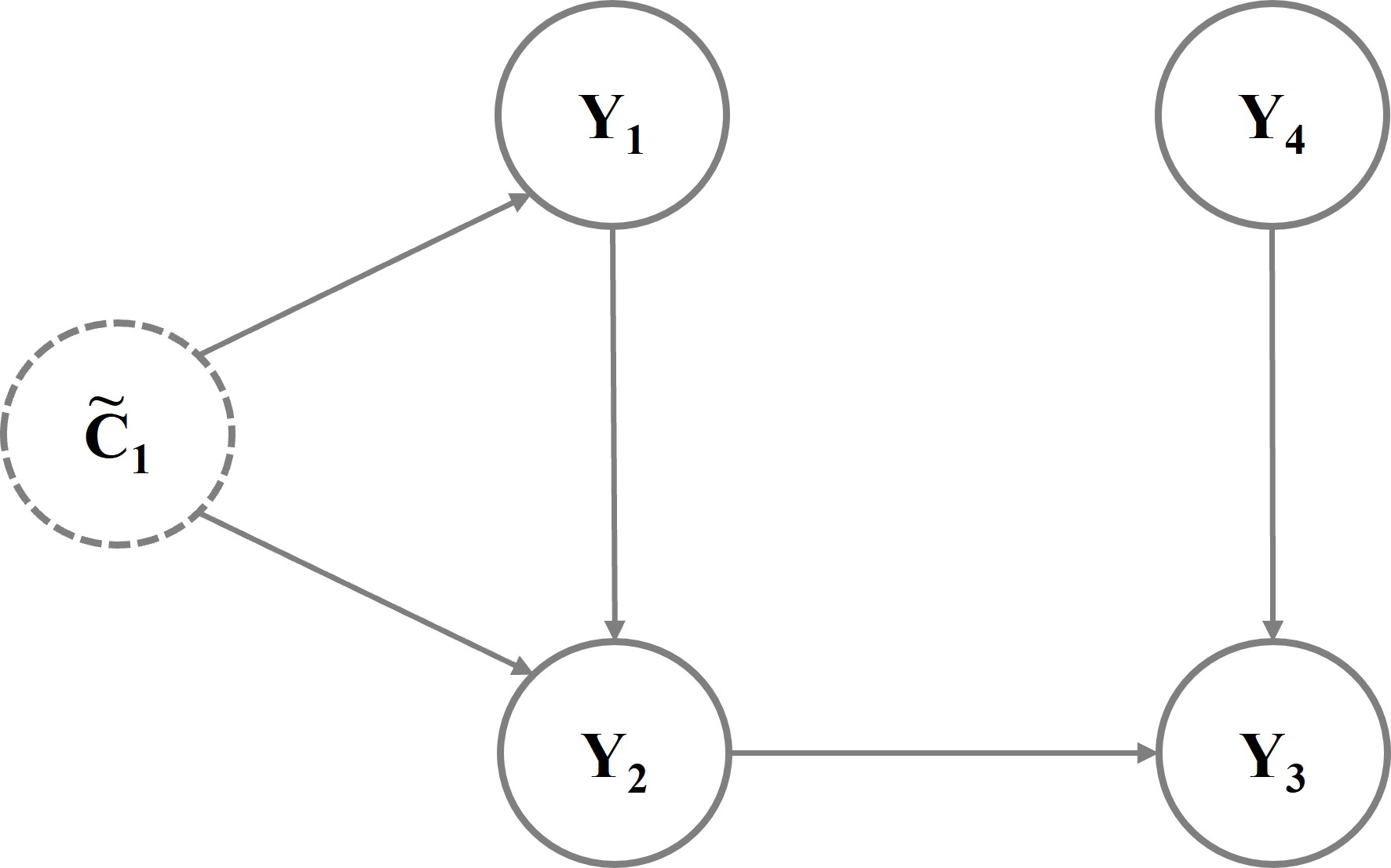}
      (b) Canonical form
  \end{subfigure}
  \caption{Graphical examples of a non-canonical form and a canonical form.}
  \label{cononical}
\end{figure}

\begin{figure}[!htb]
    \centering
      \begin{subfigure}[b]{0.45\textwidth}
      \centering
      \includegraphics[width=\textwidth]{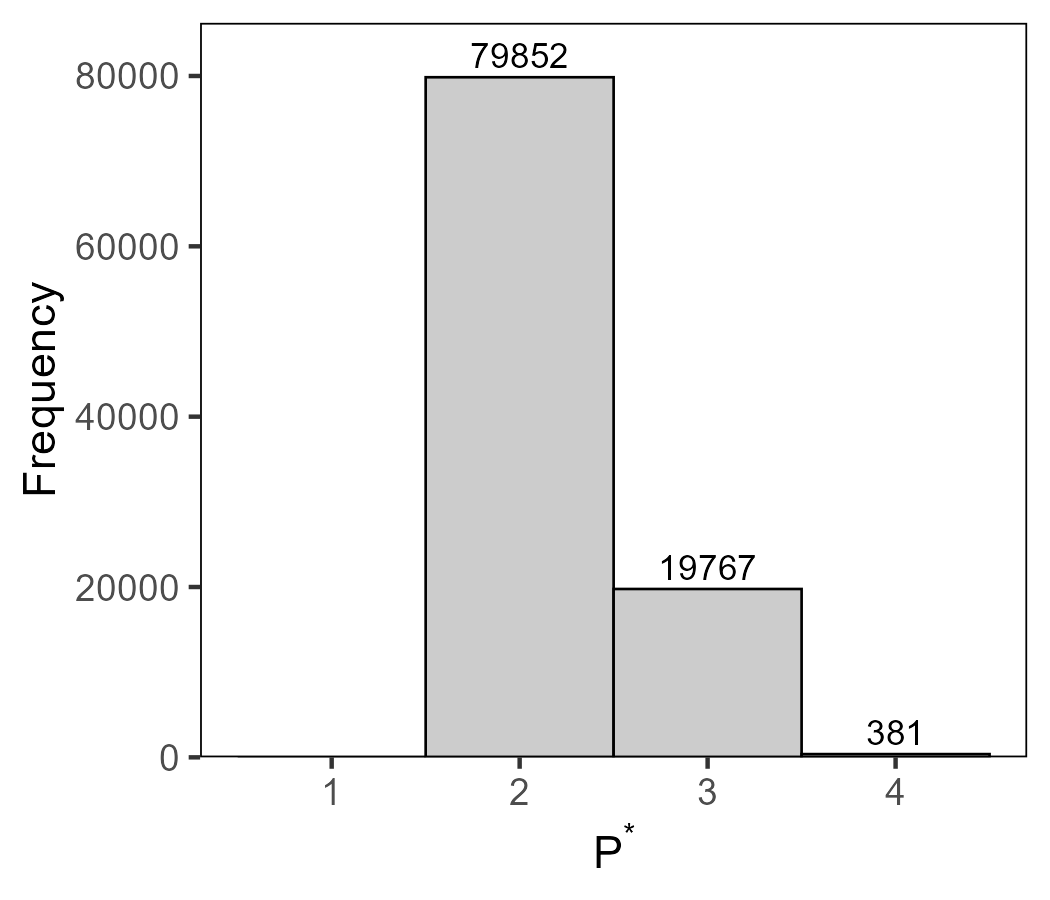}
  \end{subfigure}
  \hfill
\begin{subfigure}[b]{0.45\textwidth}
    \centering
    \includegraphics[width=\linewidth]{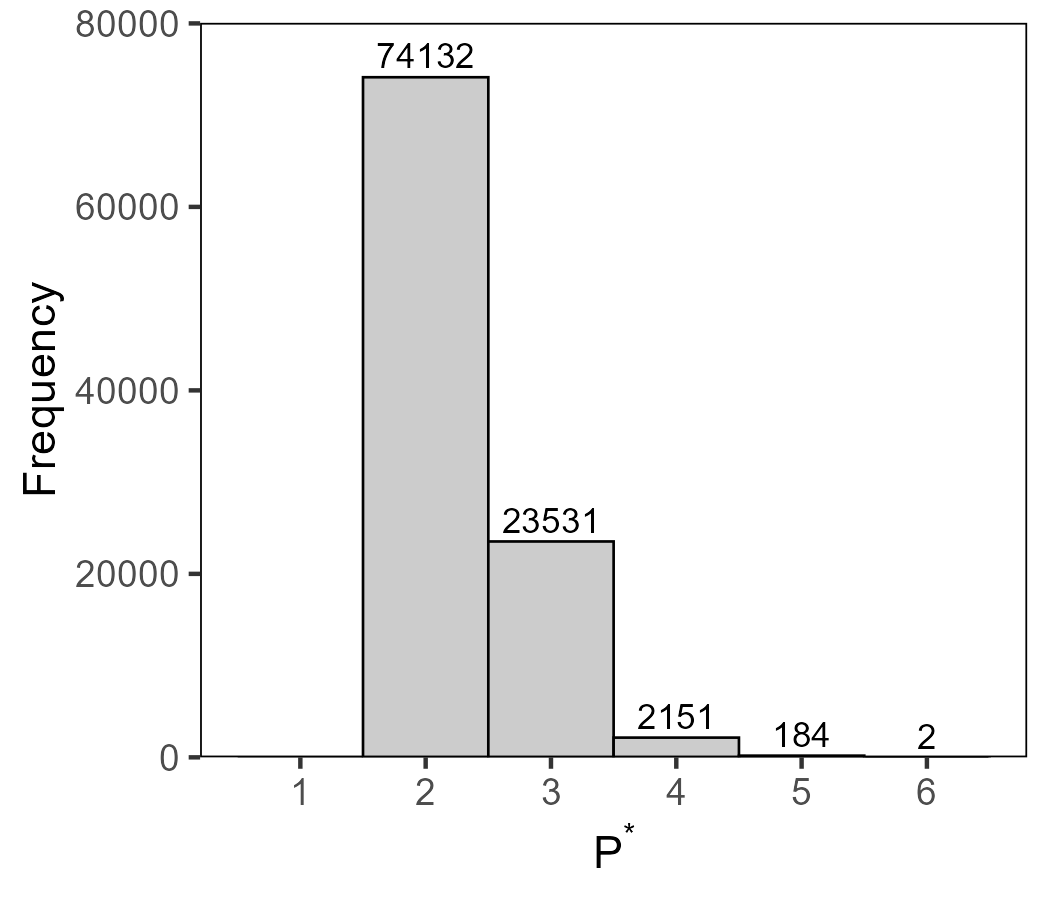}
    \end{subfigure}
    \caption{Histograms of the estimated number of latent confounders of BayCausal across all posterior samples for Scenario I (left) and Scenario II (right), aggregated over 50 repeated experiments. The simulated true number of latent confounders was set to $P^{*}=2$ in both Scenario I and II.}
\end{figure}

\end{document}